\documentclass[twocolumn]{aastex631}
\usepackage[T1]{fontenc}
\usepackage{amsmath,amssymb,bm,booktabs,graphicx}
\shorttitle{Scaled three-dimensional accretion}
\shortauthors{Mosqueira}

\newsavebox{\pdsequationbox}
\newcommand{\fitmath}[1]{%
  \sbox{\pdsequationbox}{$\displaystyle #1$}%
  \ifdim\wd\pdsequationbox>0.93\columnwidth
    \typeout{PDS equation \theequation: width \the\wd\pdsequationbox}%
    \resizebox{0.93\columnwidth}{!}{\usebox{\pdsequationbox}}%
  \else\usebox{\pdsequationbox}\fi}
\hypersetup{colorlinks=true,linkcolor=blue,citecolor=blue,urlcolor=blue,pdftitle={Scaled three-dimensional simulations of giant-planet accretion and disk clearing},pdfauthor={I. Mosqueira}}

\usepackage{placeins}

\AtBeginDocument{\raggedbottom}


\begin{document}
\title{Scaled three-dimensional simulations of giant-planet accretion and disk clearing}
\author{I. Mosqueira}
\affiliation{Department of Physics and Astronomy, San Jos\'e State University, 1 Washington Square, San Jos\'e, CA 95192, USA}
\begin{abstract}
We develop accelerated three-dimensional simulations of growing planet pairs that draw on a finite gas reservoir. An annulus compression (Scaling A) shortens the predicted clearing time and shrinks the computational domain at equal grid resolution while preserving key physical parameters. Planetary seeds start with Hill radii equal to the disk scale height, retaining three-dimensional accretion. We demonstrate the method in scaled PDS 70 and Solar-System configurations, assuming perfect Hill capture by growing planets on circular orbits. We compare Scaling A at radius-compression factor \(\Gamma=1.75\) with the unscaled PDS system at matched resolution, through a 1\% reduction of the interplanetary gas inventory. The scaled run is nearly an order of magnitude faster, with similar accreted mass and a comparable partition of gas between the planets. For the Solar-System configuration at \(\Gamma=3\), the scaling forecasts a speedup of nearly two orders of magnitude at equal local resolution. The scaling is complementary to orbital advection. On eight CPU cores of a desktop computer, the scaled PDS 70 and Solar-System runs reduce the gas inventory between the orbiting planets by roughly half in less than a day. Accretion-driven orbital convergence narrows the shared feeding zone and, together with gas clearing, limits the planetary growth budget. In our scaled PDS 70 and Solar-System runs, the angular momentum carried by meridional inflow directly constrains delivery theory. These results establish a practical route to testing accretion theory against simulations at matched scaled parameters, before applying it to unscaled planetary systems.
\end{abstract}
\section{Introduction}\label{sec:1}

Two gap-opening accreting planets share both a gas reservoir and its angular-momentum budget. Gas may be captured, pass a planet, or be displaced out of the feeding region \citep{ref29,ref30}. The angular momentum carried along each of these paths affects migration, and migration changes the gas later available to each planet. Coupled planetary accretion can produce both competition for a limited gas budget and replenishment of a planet's feeding zone \citep{ref44,ref31}. Final masses therefore depend on angular-momentum delivery as well as on the total accretion rate. In this paper we develop a scaling that makes simulations of this coupled problem affordable, apply it to scaled PDS 70 and Solar-System planet pairs, and measure the mass and angular momentum delivered to each planet. In future work we will use these measurements to develop and test a theory of mass and angular-momentum delivery that links planetary growth, migration, and the evolving gas supply \citep{ref28}.

Besides the Jupiter--Saturn system, the PDS 70 observations establish an analogous coupled architecture: two accreting giant planets occupy a common gap \citep{ref56,ref57,ref36}, and a circumplanetary disk is detected around the outer planet \citep{ref58,ref37}. \citet{ref32} connects this observed gap-opening pair to its circumplanetary supply and delivery history. A subsequent analysis of PDS 70 c and SR 12 c \citep{ref6} supplies the predictive closure we adopt here, in which planets draw on a finite interplanetary mass reservoir in a low-viscosity disk. In this setting accretion competes with torque-driven gas clearing, limiting further growth.

The present inviscid, finite-reservoir simulations implement that framework. Throughout our model and results, ``planet'' denotes the composite of the seed and its retained circumplanetary gas; planetary masses include both. The simulations follow how the shared supply is divided between the two growing planets and the gas displaced beyond their orbits, retaining the orbital and internal angular momentum of captured material. Accretion contributes substantially to orbital evolution (Section~\ref{sec:5.5}) and thereby changes the reservoir later accessible to each planet. Predicting the mass partition and predicting angular-momentum delivery are therefore parts of the same closure. The scaling developed below makes this coupled evolution affordable to simulate.

Meridional flows around gap-opening planets were first identified in the global three-dimensional simulations of \citet{Kley2001}. Distance from the midplane affects the routes by which gas reaches the Hill region and the angular momentum it carries. This motivates measurements of joint mass and angular-momentum fluxes \citep{ref10,ref14,ref11}. Later studies further characterize the intermediate-latitude inflow and near-midplane outflow \citep{ref13}. Related numerical work addresses simultaneous gas accretion by two planets \citep{ref21}, consistent accretion torques during planetary migration \citep{ref18,ref19}, and the PDS 70 pair itself \citep{ref59,ref60}.

This study advances two scalings, Scaling A and Scaling B, intended to make coupled planetary growth and disk clearing computationally affordable (see \citet{ref5} for the precedent). Our strategy is to identify the physical quantities the scaled system must retain. Scaling A contracts the radial domain as well as the orbital radii, while preserving the local Hill, thermal, and orbital-frequency scales. This physical parameter transformation is distinct from orbital advection, which accelerates the numerical evolution of a specified system \citep{ref33,ref55}. As we show below, the two approaches are complementary.

We focus here on all-radii full-annulus compression (Scaling A) and leave further evaluation of azimuthal periodic-sector compression (Scaling B) for later work. We apply Scaling A to PDS 70 and Solar-System configurations and measure its computational advantage through a matched comparison with an unscaled simulation.
The paper is organized as follows. Section~\ref{sec:2} develops Scaling A and Scaling B, the torque-budget clearing time, the separate equal-resolution grid saving, and the limits on compression; Section~\ref{sec:3} applies Scaling A to scaled PDS 70 and Solar-System configurations. Section~\ref{sec:4} describes the numerical model and delivery diagnostics and presents the matched benchmark and the extended simulations, and Section~\ref{sec:5} analyzes the angular-momentum budgets behind convergent migration. Section~\ref{sec:6} explains how these measurements will test the delivery theory developed in the companion paper, and Section~\ref{sec:7} summarizes the conclusions.

\section{Scaling at a fixed starting regime}\label{sec:2}

\subsection{Preserving the three-dimensional accretion regime}\label{sec:2.1}

Let \(a_i\) and \(M_i\) be the orbital radius and total mass of planet \(i\). Throughout the paper, \(i=b,c\) labels the inner and outer planets, respectively, in either planetary configuration. We use star-centered cylindrical radius \(R\), height \(z\), and azimuth \(\phi\). Spherical radius is \(r=(R^2+z^2)^{1/2}\), and polar angle \(\theta\) is measured from the positive \(z\) axis, so \(R=r\sin\theta\) and \(z=r\cos\theta\). The disk midplane is \(z=0\), where \(r=R\); the positive \(z\) axis sets the sign of angular momentum. Gas volume density is denoted by \(\rho\). Let \(G\) be the gravitational constant, \(M_\star\) the central stellar mass, and \(c_s\) the constant isothermal sound speed. The Keplerian orbital frequency \(\Omega_i\), gas scale height \(H_i\), and Hill radius \(R_{H,i}\) are defined by \begin{equation}\label{eq:1}
\fitmath{\begin{gathered}
\Omega_i^2=\frac{GM_\star}{a_i^3},\\[3pt]
 H_i=\frac{c_s}{\Omega_i},\\[3pt]
 R_{H,i}^3=\frac{GM_i}{3\Omega_i^2}.
\end{gathered}}
\end{equation} Define the aspect ratio \(h_i=H_i/a_i\), planet-to-star mass ratio \(q_i=M_i/M_\star\), thermal-mass parameter \(q_{{\rm th},i}=q_i/h_i^3\), and Bondi radius \(R_{B,i}=GM_i/c_s^2\). We choose the initial masses so that \(R_{H,i}=H_i\). Equivalently, \begin{equation}\label{eq:2}
\fitmath{M_i=3M_\star h_i^3,\qquad
 q_{{\rm th},i}=3,\qquad
 \frac{R_{B,i}}{H_i}=3.}
\end{equation} The capture region is therefore Hill-limited and comparable to the vertical gas thickness. This is the regime we wish to retain when designing faster simulations.\footnote{``Fixed \(R_H/H\)'' describes the comparison of initial scaled configurations, not a constraint imposed throughout accretion. Masses, Hill radii, and orbits subsequently evolve. The initial state is also distinct from the strongly superthermal asymptote of some accretion fits \citep{ref2}.} At constant sound speed, the seed masses vary as \(a_i^{3/2}\); the two planets need not have equal masses.

\subsection{Scaling A: all-radii full-annulus compression}\label{sec:2.2}

Unprimed quantities refer to the reference system; primes denote the scaled system. The radial mapping below is written in the midplane, where \(r=R\). Let \(\Sigma(r)\) be gas surface density and \(M_g\) the total gas mass represented in the domain.

Let \(\Gamma>1\) denote the radius-compression factor, the reference radius divided by its scaled value. Take \begin{equation}\label{eq:3}
\fitmath{\begin{aligned}
 a_i'&=a_i/\Gamma,& r'&=r/\Gamma,\\
 M_\star'&=M_\star/\Gamma^3,& M_i'&=M_i,\\
 c_s'&=c_s,& M_g'&=M_g.
 \end{aligned}}
\end{equation} For the homologously contracted surface-density profile, preservation of represented gas mass gives \(\Sigma'(r/\Gamma)=\Gamma^2\Sigma(r)\).

\textbf{Preserved local physics.} Substitution into the definitions in Section~\ref{sec:2.1} gives \(\Omega_i'=\Omega_i\): the reduction in central mass exactly compensates the contraction of orbital radius. The Keplerian orbital period \(P_i=2\pi/\Omega_i\), physical scale height \(H_i\), Hill radius \(R_{H,i}\), and Bondi radius \(R_{B,i}\) are therefore unchanged. So are the sound-crossing times \(H_i/c_s\) and \(R_{H,i}/c_s\), the ratios \(R_H/H\) and \(R_B/H\), and the thermal-mass parameter \(q_{\rm th}\). The planet masses, their ratio, total represented gas mass, and gas-to-seed mass ratios are retained. Temperature is unchanged at fixed gas composition.

The force balance explains why these invariants matter. In the local frame rotating with planet \(i\), let \(x\) be radial displacement, \(z\) height above the midplane, and \(u_y\) the background azimuthal velocity relative to the planet's circular motion. To leading order in distance from the planet divided by its orbital radius, the stellar tidal and centrifugal terms and the Keplerian shear are \begin{equation}\label{eq:4}
\fitmath{\begin{gathered}
a_{x,\rm tide}=3\Omega_i^2x,\\[3pt]
 a_{z,\rm tide}=-\Omega_i^2z,\\[3pt]
 u_y=-\frac32\Omega_i x.
\end{gathered}}
\end{equation} At fixed physical \(x\) and \(z\), all three expressions are unchanged; the Coriolis coefficient \(2\Omega_i\) is also unchanged. At the same planet-centered distance the planetary gravity is unchanged because \(M_i\) is retained, provided any gravitational softening is kept at the same fraction of the preserved local scales. The thermal acceleration scale \(c_s^2/H_i\) is likewise unchanged. Thus the scaling retains the leading local competition between planetary gravity, stellar tides, shear, rotation, and pressure that governs Hill-scale capture. This is a local force-scale statement, not an assertion that the full curved geometry or the evolved pressure gradients are identical. The local normalization has a precedent in \citet[\S{}2.1]{ref24}, who use \(H\) and \(\Omega^{-1}\) as length and time units for non-self-gravitating isothermal flow, with planetary mass parameter \(R_B/H=q_{\rm th}\). In our notation this also equals \(3(R_H/H)^3\). Scaling A preserves these equivalent local mass parameters at both planets while changing the global reservoir geometry.

The corresponding local velocity and planet-centered specific-angular-momentum scales, denoted \(v_{H,i}\) and \(j_{H,i}\), are \begin{equation}\label{eq:5}
\fitmath{v_{H,i}=\Omega_iR_{H,i},\qquad j_{H,i}=\Omega_iR_{H,i}^2.}
\end{equation} Both are invariant. In particular, the ratio \(v_{H,i}/c_s=R_{H,i}/H_i\) is retained, and the circularization radius associated with the angular-momentum scale is \(j_{H,i}^2/(GM_i)=R_{H,i}/3\). Section~\ref{sec:4.3} uses this unchanged normalization to measure the actual inflow angular momentum. The measured inflow need not equal \(j_{H,i}\): its magnitude, sign, and distribution among streams are quantities the simulations must determine.

\textbf{Quantities that change.} The aspect ratios \(H/a\) and \(R_H/a\) increase by \(\Gamma\), while planetary separation in Hill radii decreases by \(\Gamma\). The planet-to-star mass ratio increases by \(\Gamma^3\). The Keplerian orbital speed \(v_{K,i}=a_i\Omega_i\) decreases by \(\Gamma\), and the star-centered circular specific angular momentum \(\ell_{K,i}=a_i^2\Omega_i\) decreases by \(\Gamma^2\). This orbital angular-momentum scale is distinct from the preserved internal scale \(j_{H,i}\). Surface density increases by \(\Gamma^2\), so local gas mass loading is not invariant even though the total reservoir mass is. For a column with the same normalized vertical profile and physical \(H\), its volume-density and isothermal-pressure normalizations also increase by \(\Gamma^2\). These changes supply the faster delivery and clearing discussed below. Preservation under the parameter transformation is distinct from conservation of mass and angular momentum during each simulation; it does not require the two systems to have the same total angular momentum.

The organizing idea of preserving selected dynamical balances is motivated by the ring scaling of \citet{ref5}. The gas transformation above is our construction, not a direct application of their prescription.

\subsection{Scaling B: periodic-sector compression}\label{sec:2.3}

Let \(S>1\) be the azimuthal compression factor and \(L_\phi\) the domain's angular extent in radians, equal to \(2\pi\) in the full-annulus reference. Keep the central mass, radial geometry, planet masses, and sound speed unchanged, but set \begin{equation}\label{eq:6}
\fitmath{L_\phi'=\frac{2\pi}{S},\qquad \Sigma'=S\Sigma.}
\end{equation} The density normalization preserves the gas mass actually represented in the sector. Local orbital and capture scales, including \(H/a\) and \(R_H/a\), remain unchanged. The local times, force coefficients, \(v_H\), and \(j_H\) just described are also retained; unlike Scaling A, the circular orbital speed and star-centered specific angular momentum do not change. For a matched vertical profile the density and pressure normalizations increase by \(S\). The shorter circumference instead changes encounter frequency and the permitted global azimuthal structure. The theory and fluid calculation must use the same periodic geometry. No manual multiplier is applied to Athena++'s resolved planetary forces.

This altered encounter history is particularly relevant to capture versus passage. At a fixed gas--planet relative angular speed, the interval between successive periodic encounters is shorter by a factor \(S\). Gas that passes a planet without capture can therefore meet its periodic image sooner, with less time for intervening radial and vertical transport. Repeated encounters can change the fraction ultimately captured rather than passed onward, as well as the division of supply between the two planets. Preserving the local capture scales does not preserve this history, so the change need not be merely a faster version of the same delivery process. This is a physical reason for prioritizing Scaling A in the present paper: it retains the full azimuth and does not introduce these additional periodic repetitions. Detailed evaluation of periodic-sector compression, including its capture-to-passage partition in the periodic geometry, is left to a later paper; its nominal scalings below are given for context.

\subsection{Clearing time from the torque budget}\label{sec:2.4}

This calculation estimates physical gas clearing with the planetary radii and reservoir boundaries held fixed, so it excludes the decrease in interplanetary inventory caused by the planets migrating toward one another. The simulations instead measure the inventory between the moving orbits (Section~\ref{sec:4.5}), whose reduction includes both effects.

The relevant time is the angular momentum needed to displace the gas divided by the torque supplied by the planets, as in the earlier gap-opening estimate of \citet{ref44}. The tidal exchange is mediated by planet-excited density waves \citep{ref46}. Following the finite-reservoir estimate in \citet[\S{}IV]{ref6}, use a clearing distance \(w_i=\eta_wR_{H,i}\), where \(\eta_w\) is a fixed dimensionless width in Hill radii. Let \(\Delta\ell_i\) be the required change in stellar specific angular momentum, \(\mathcal T_i\) the clearing torque exerted by planet \(i\), and \(\Sigma_{\rm gap}\) the mean surface density of the shared reservoir. Define \(K_i\) as the torque magnitude per unit surface density, with dimensionless coefficient \(C_T\). The estimates are \begin{equation}\label{eq:7}
\fitmath{\begin{gathered}
\Delta\ell_i\simeq\frac12a_i\Omega_i w_i,\\[3pt]
 |\mathcal T_i|=K_i\Sigma_{\rm gap},\\[3pt]
 K_i=C_Tq_i^2a_i^4\Omega_i^2h_i^{-3}.
\end{gathered}}
\end{equation} The clearing lobes are the inner-directed torque of the inner planet and the outer-directed torque of the outer planet. Other torque contributions redistribute gas within the shared reservoir. Using the mean interplanetary surface density to normalize these exterior clearing lobes is a closure: it assumes that transport across the planetary orbits supplies those lobes in proportion to the shared inventory. The coefficient \(C_T\) and width \(\eta_w\) summarize torque delivery and the displacement required for removal. Their fixed values are the assumptions behind the scaling estimate, not measurements of that transport. A torque impulse alone need not permanently remove gas from the feeding region; capture, storage, and return flows are followed by the simulation. Wave excitation and deposition are distinct: nonlinear steepening and shock damping determine where the wave angular momentum reaches the gas \citep{ref47,ref48}. That weakly nonlinear propagation theory separates excitation from deposition, but it is not a calibrated damping law for the superthermal planets evolved here.

Let \(M_{\rm gap}\) and \(A_{\rm gap}\) denote the reservoir mass and area, and \(\dot M_{\rm clear}>0\) the rate of torque-driven removal. Substitute \(\Sigma_{\rm gap}=M_{\rm gap}/A_{\rm gap}\) and \(\dot M_{\rm clear}=\sum_i|\mathcal T_i|/\Delta\ell_i\). The resulting torque-only half-clearing time, denoted \(t_{1/2,T}\), is \begin{equation}\label{eq:8}
\fitmath{t_{1/2,T}=\frac{(\ln2)A_{\rm gap}}
                    {\displaystyle\sum_i K_i/\Delta\ell_i}.}
\end{equation} Here \(A_{\rm gap}\), the orbital radii, and the coefficients \(K_i/\Delta\ell_i\) remain fixed during the estimate. Only torque-driven gas removal decreases \(M_{\rm gap}\); neither capture nor migration-induced reclassification is included in \(t_{1/2,T}\). More generally, for a removed fraction \(0<f<1\), the same fixed-coefficient estimate gives \begin{equation}\label{eq:9}
\fitmath{t_{f,T}=\frac{-\ln(1-f)A_{\rm gap}}
                   {\displaystyle\sum_i K_i/\Delta\ell_i}.}
\end{equation} Choosing 1\% rather than half-clearing replaces \(\ln2\) with \(\ln(100/99)\), without changing the predicted compression dependence. This generalization concerns the torque-only estimate; the measured moving-orbit inventory benchmark defined in Section~\ref{sec:4.5} includes capture and migration as well. At the chosen \(R_H=H\) starting state, this simplifies to \begin{equation}\label{eq:10}
\fitmath{t_{1/2,T}=
 \frac{\eta_w(\ln2)A_{\rm gap}}
      {18C_T\displaystyle\sum_i\Omega_iH_i^2}.}
\end{equation} The preserved local scales leave the denominator unchanged. The physical acceleration is therefore set by the reduction in represented reservoir area: \begin{equation}\label{eq:11}
\fitmath{\frac{t_{1/2,T}'}{t_{1/2,T}}=
 \begin{cases}
  \Gamma^{-2},&\text{A: all radii},\\
  S^{-1},&\text{periodic sector}.
 \end{cases}}
\end{equation} For all-radii compression, the same reservoir mass experiences a torque larger by \(\Gamma\), while the required angular-momentum transfer is smaller by \(\Gamma\). These two physical effects give the \(\Gamma^2\) acceleration. No grid argument enters this result.

This estimate holds the torque coefficient and Hill-scaled clearing width fixed. It does not contract that local displacement with the global annulus. A net Hill-capture law proportional to \(\Sigma\Omega R_H^2\) adds a loss channel with the same area scaling at fixed coefficients. Because the simulations evolve masses, orbits, profiles, and delivery, their moving-orbit inventory reduction is not a direct measurement of \(t_{1/2,T}\). A like-for-like comparison uses fixed-initial-orbit inventories, separates capture from torque-driven removal, and reports the migration contribution separately. The \(\Gamma^2\) scaling therefore applies to this fixed-orbit torque budget, but is not established for the combined clearing-and-migration diagnostic.

\subsection{Equal-resolution grid savings and total computational gain}\label{sec:2.5}

Compression also reduces the computational domain. Because the physical \(H\) and \(R_H\) are preserved, the local cell widths can be retained. In the all-radii construction, radial and azimuthal extents each shrink by \(\Gamma\), while the represented physical vertical thickness stays fixed. The nominal cell count consequently decreases by \(\Gamma^2\). Sector compression reduces only the azimuthal angle. Let \(N\) and \(N'\) be the reference and scaled cell counts at equal physical resolution. Define the grid-saving factor \(\mathcal S_{\rm grid}\) as their ratio: \begin{equation}\label{eq:12}
\fitmath{\mathcal S_{\rm grid}\equiv\frac{N}{N'}=
 \begin{cases}
 \Gamma^2,&\text{A: all radii},\\
 S,&\text{periodic sector}.
 \end{cases}}
\end{equation}

This is a smaller mesh at equal physical resolution, not coarsening. Let \(T_{\rm wall}\) and \(T_{\rm wall}'\) be the reference and scaled execution times to a common fractional inventory reduction. At unchanged timestep and cell-proportional update cost, the total computational gain, denoted \(\mathcal S_{\rm wall}\), is the reference runtime divided by the scaled runtime. It is the product of the physical acceleration and the equal-resolution grid saving: \begin{equation}\label{eq:13}
\fitmath{\mathcal S_{\rm wall}\equiv\frac{T_{\rm wall}}{T_{\rm wall}'}\simeq
 \begin{cases}
 \Gamma^2\times\Gamma^2=\Gamma^4,&\text{A: all radii},\\
 S\times S=S^2,&\text{periodic sector}.
 \end{cases}}
\end{equation} Thus the full-annulus all-radii construction provides a fourth-power nominal computational benefit: a second-power reduction in physical clearing time and a separate second-power reduction in cells at the same resolution. No larger timestep or coarser cells are assumed. For example, \(\Gamma=1.75\) gives a nominal combined computational gain of \(1.75^4\simeq9.38\): \(3.06\) from faster physical half-clearing multiplied by \(3.06\) from the smaller, equally resolved mesh. The overall benefit is therefore approximately ninefold, not threefold. Local capture work, refinement coverage, and output can reduce the cell-count benefit; any timestep change is an additional measured factor.

\textbf{Velocity-ceiling contribution to the timestep.} At equal physical cell width and unchanged sound speed, an acoustic limiter introduces no systematic compression penalty. The numerical velocity ceiling can instead provide an additional timestep gain. Following the component-wise, sign-preserving operation of \citet[\S{}2.4]{ref34}, we clip velocity components and update their stored momenta. With \(r_{\rm inner}\) denoting the inner radial boundary, the ceiling is \begin{equation}\label{eq:14}
\fitmath{\begin{gathered}
v_{\rm cap}=3\sqrt{2GM_\star/r_{\rm inner}},\\[3pt]
 v_{\rm cap}'=v_{\rm cap}/\Gamma.
\end{gathered}}
\end{equation} The primed relation applies to Scaling A only. We adopt a coefficient of three rather than their ten, and retain the fixed isothermal sound speed. \citet[\S{}3.4.2]{ref35} provide a related disk-atmosphere precedent, but use a density-dependent ceiling on radial and polar motion rather than our global component-wise prescription. Both benchmark runs use the same coefficient and algorithm.

For corresponding ceiling-controlled cells at matched physical width, the local timestep ratio, denoted \(\chi_{\rm ceil,loc}\), is approximately \begin{equation}\label{eq:15}
\fitmath{\chi_{\rm ceil,loc}\simeq
 \frac{v_{\rm cap}+c_s}{v_{\rm cap}/\Gamma+c_s}.}
\end{equation} It approaches \(\Gamma\) when velocity dominates and unity when sound speed dominates. This is a contribution from the numerical prescription, not a scaling law for atmospheric flow. Clipping changes gas momentum; its contribution to the budget is distinguished in Section~\ref{sec:4.2} and Section~\ref{sec:5.1}. The limiting cell can change during evolution. Appendix~\ref{sec:A} gives the Courant--Friedrichs--Lewy (CFL) timestep diagnostics and the sampled below-ceiling and ceiling-limited states.

\textbf{Measured three-factor benchmark.} The common endpoint is a 1\% reduction of post-initial-capture interplanetary inventory (Section~\ref{sec:4.5}). The notation below describes Scaling A against the uncompressed reference. Let \(t_{1,\Gamma}\), \(\mathcal N_{1,\Gamma}\), and \(T_{{\rm wall},1}(\Gamma)\) be the physical duration, accepted step count, and accumulated execution time to that endpoint. Define \begin{equation}\label{eq:16}
\fitmath{\begin{aligned}
 \overline{\Delta t}_{1,\Gamma}&=t_{1,\Gamma}/\mathcal N_{1,\Gamma},\\
 \bar c_{{\rm step},1,\Gamma}&=T_{{\rm wall},1}(\Gamma)/\mathcal N_{1,\Gamma},\\
 \chi_{1}&=\overline{\Delta t}_{1,\Gamma}/\overline{\Delta t}_{1,1}.
 \end{aligned}}
\end{equation} The mean step cost includes coupling trials, segment startup, and I/O under the same timing convention. The speedup then satisfies the identity \begin{equation}\label{eq:17}
\fitmath{\mathcal S_{1}=
 \underbrace{\frac{t_{1,1}}{t_{1,\Gamma}}}_{\text{duration}}
 \underbrace{\frac{\bar c_{{\rm step},1,1}}
 {\bar c_{{\rm step},1,\Gamma}}}_{\text{step cost}}
 \underbrace{\chi_{1}}_{\text{mean timestep}}.}
\end{equation} The comparison measures each factor through the same inventory target; no timestep scaling is assumed in this identity. For Scaling A, under the nominal torque-duration and cell-proportional cost estimates, it becomes \(\mathcal S_{1}\simeq\Gamma^4\chi_{1}\). In the ceiling-dominated limit \(\chi_{1}\simeq\Gamma\) it would give \(\Gamma^5\), a conditional estimate rather than a measured decomposition. At \(\Gamma=1.75\), the equal-timestep and ceiling-dominated estimates are 9.4 and 16.4, respectively; these are alternatives, not factors to multiply again. Section~\ref{sec:4.5} reports the measured 9.31-fold saving and its three factors. Scaling B supplies no analogous systematic ceiling contribution.

\textbf{Effective compressions.} The nominal \(\Gamma\) describes the physical transformation; a particular mesh and solver need not realize each factor at that value. We therefore define effective compressions from the measured duration and from the step-weighted mean cell count \(\langle N\rangle_\Gamma\) over the run: \begin{equation}\label{eq:effcomp}
\fitmath{\Gamma_{\rm dur}\equiv\left(\frac{t_{1,1}}{t_{1,\Gamma}}\right)^{1/2},\qquad
 \Gamma_{\rm grid}\equiv\left(\frac{\langle N\rangle_{1}}{\langle N\rangle_{\Gamma}}\right)^{1/2},}
\end{equation} together with the ratio of cost per cell-step, \begin{equation}\label{eq:kappa}
\fitmath{\eta_{\rm cost}\equiv\frac{\bar c_{{\rm step},1,1}/\langle N\rangle_{1}}
 {\bar c_{{\rm step},1,\Gamma}/\langle N\rangle_{\Gamma}}.}
\end{equation} The step-cost factor is then \(\Gamma_{\rm grid}^2\eta_{\rm cost}\), and the identity becomes \begin{equation}\label{eq:s1eff}
\fitmath{\mathcal S_{1}=\Gamma_{\rm dur}^2\,\Gamma_{\rm grid}^2\,\eta_{\rm cost}\,\chi_{1}.}
\end{equation} Nominally \(\Gamma_{\rm dur}=\Gamma_{\rm grid}=\Gamma\) and \(\eta_{\rm cost}=1\), which recovers \(\Gamma^4\chi_{1}\). The effective grid compression falls below \(\Gamma\) when refined regions of fixed physical size, such as those around the planets, occupy a significant share of the mesh; \(\eta_{\rm cost}\) falls below unity when work that does not scale with cell count, such as coupling trials and capture-patch updates, is significant.

A deliberately coarser exploratory grid is a further choice, not part of these scaling gains. We report the realized physical resolution separately from compression; a resolution-sensitivity sequence has not been completed.

\subsection{Scaling parameters at a glance}\label{sec:2.6}

Table~\ref{tab:1} summarizes ratios of scaled to reference quantities, including the resulting computational time. Planet masses, sound speed, and represented reservoir mass are held fixed. The stellar-Keplerian Toomre estimate is \(Q_{\rm K}=c_s\Omega_K/(\pi G\Sigma)\), with \(\Omega_K=(GM_\star/R^3)^{1/2}\); Section~\ref{sec:2.7} gives its interpretation.

\begin{table}[tp]
\centering
\caption{Parameters for Scaling A (all-radii compression) and Scaling B (periodic-sector compression). Entries are scaled-to-reference ratios except for the orbital maps. Computational time assumes unchanged timestep and cell-proportional cost.}\label{tab:1}
\small\setlength{\tabcolsep}{2pt}
\begin{tabular}{@{}lll@{}}
\toprule
\parbox[t]{0.4800\columnwidth}{\raggedright Quantity ratio or map\strut} & \parbox[t]{0.2400\columnwidth}{\centering Scaling A: all radii\strut} & \parbox[t]{0.2400\columnwidth}{\centering Scaling B: periodic sector\strut} \\[4pt]
\midrule
\parbox[t]{0.4800\columnwidth}{\raggedright Central mass\strut} & \parbox[t]{0.2400\columnwidth}{\centering \(\Gamma^{-3}\)\strut} & \parbox[t]{0.2400\columnwidth}{\centering \(1\)\strut} \\[4pt]
\parbox[t]{0.4800\columnwidth}{\raggedright Orbital-radius map\strut} & \parbox[t]{0.2400\columnwidth}{\centering \(a_i'=a_i/\Gamma\)\strut} & \parbox[t]{0.2400\columnwidth}{\centering \(a_i'=a_i\)\strut} \\[4pt]
\parbox[t]{0.4800\columnwidth}{\raggedright Radial width and planetary separation\strut} & \parbox[t]{0.2400\columnwidth}{\centering \(\Gamma^{-1}\)\strut} & \parbox[t]{0.2400\columnwidth}{\centering \(1\)\strut} \\[4pt]
\parbox[t]{0.4800\columnwidth}{\raggedright Azimuthal angle\strut} & \parbox[t]{0.2400\columnwidth}{\centering \(1\)\strut} & \parbox[t]{0.2400\columnwidth}{\centering \(S^{-1}\)\strut} \\[4pt]
\parbox[t]{0.4800\columnwidth}{\raggedright Surface density\strut} & \parbox[t]{0.2400\columnwidth}{\centering \(\Gamma^2\)\strut} & \parbox[t]{0.2400\columnwidth}{\centering \(S\)\strut} \\[4pt]
\parbox[t]{0.4800\columnwidth}{\raggedright Stellar-Keplerian Toomre \(Q_{\rm K}\)\strut} & \parbox[t]{0.2400\columnwidth}{\centering \(\Gamma^{-2}\)\strut} & \parbox[t]{0.2400\columnwidth}{\centering \(S^{-1}\)\strut} \\[4pt]
\parbox[t]{0.4800\columnwidth}{\raggedright Planet and represented gas masses\strut} & \parbox[t]{0.2400\columnwidth}{\centering \(1\)\strut} & \parbox[t]{0.2400\columnwidth}{\centering \(1\)\strut} \\[4pt]
\parbox[t]{0.4800\columnwidth}{\raggedright Sound speed and temperature\strut} & \parbox[t]{0.2400\columnwidth}{\centering \(1\)\strut} & \parbox[t]{0.2400\columnwidth}{\centering \(1\)\strut} \\[4pt]
\parbox[t]{0.4800\columnwidth}{\raggedright \(\Omega,\ H,\ R_H\)\strut} & \parbox[t]{0.2400\columnwidth}{\centering \(1\)\strut} & \parbox[t]{0.2400\columnwidth}{\centering \(1\)\strut} \\[4pt]
\parbox[t]{0.4800\columnwidth}{\raggedright Bondi radius \(R_B\)\strut} & \parbox[t]{0.2400\columnwidth}{\centering \(1\)\strut} & \parbox[t]{0.2400\columnwidth}{\centering \(1\)\strut} \\[4pt]
\parbox[t]{0.4800\columnwidth}{\raggedright \(R_H/H,\ R_B/H,\ q_{\rm th}\)\strut} & \parbox[t]{0.2400\columnwidth}{\centering \(1\)\strut} & \parbox[t]{0.2400\columnwidth}{\centering \(1\)\strut} \\[4pt]
\parbox[t]{0.4800\columnwidth}{\raggedright \(P,\ H/c_s,\ R_H/c_s\)\strut} & \parbox[t]{0.2400\columnwidth}{\centering \(1\)\strut} & \parbox[t]{0.2400\columnwidth}{\centering \(1\)\strut} \\[4pt]
\parbox[t]{0.4800\columnwidth}{\raggedright \(v_H,\ j_H\)\strut} & \parbox[t]{0.2400\columnwidth}{\centering \(1\)\strut} & \parbox[t]{0.2400\columnwidth}{\centering \(1\)\strut} \\[4pt]
\parbox[t]{0.4800\columnwidth}{\raggedright \(H/a,\ R_H/a\)\strut} & \parbox[t]{0.2400\columnwidth}{\centering \(\Gamma\)\strut} & \parbox[t]{0.2400\columnwidth}{\centering \(1\)\strut} \\[4pt]
\parbox[t]{0.4800\columnwidth}{\raggedright Circular orbital speed \(v_K\)\strut} & \parbox[t]{0.2400\columnwidth}{\centering \(\Gamma^{-1}\)\strut} & \parbox[t]{0.2400\columnwidth}{\centering \(1\)\strut} \\[4pt]
\parbox[t]{0.4800\columnwidth}{\raggedright Orbital specific angular momentum \(\ell_K\)\strut} & \parbox[t]{0.2400\columnwidth}{\centering \(\Gamma^{-2}\)\strut} & \parbox[t]{0.2400\columnwidth}{\centering \(1\)\strut} \\[4pt]
\parbox[t]{0.4800\columnwidth}{\raggedright Nominal physical half-clearing time\strut} & \parbox[t]{0.2400\columnwidth}{\centering \(\Gamma^{-2}\)\strut} & \parbox[t]{0.2400\columnwidth}{\centering \(S^{-1}\)\strut} \\[4pt]
\parbox[t]{0.4800\columnwidth}{\raggedright Cells at equal physical resolution\strut} & \parbox[t]{0.2400\columnwidth}{\centering \(\Gamma^{-2}\)\strut} & \parbox[t]{0.2400\columnwidth}{\centering \(S^{-1}\)\strut} \\[4pt]
\parbox[t]{0.4800\columnwidth}{\raggedright Nominal computational time at equal timestep\strut} & \parbox[t]{0.2400\columnwidth}{\centering \(\Gamma^{-4}\)\strut} & \parbox[t]{0.2400\columnwidth}{\centering \(S^{-2}\)\strut} \\[4pt]
\parbox[t]{0.4800\columnwidth}{\raggedright CFL ratio, acoustic / scaled-speed limits\strut} & \parbox[t]{0.2400\columnwidth}{\centering \(1\) / \(\Gamma\), conditional\strut} & \parbox[t]{0.2400\columnwidth}{\centering No intrinsic extra factor\strut} \\[4pt]
\bottomrule
\end{tabular}
\end{table}

\subsection{Limits on increasing compression}\label{sec:2.7}

The useful compression is limited by the intended evolution, not only by whether the initial conditions fit on a grid.

For Scaling A, \(M_i/M_\star\) grows as \(\Gamma^3\), while \(H/a\) and \(R_H/a\) grow as \(\Gamma\). Excessive compression weakens the central-mass and local-disk approximations. In the all-radii version, the unchanged Hill spheres occupy a larger fraction of the contracted planetary separation. Their buffered numerical patches must remain separate as masses grow and orbits migrate. Boundaries must also leave room for the flow and torque deposition.

\textbf{Toomre Q as a physical diagnostic.} Preserving \(R_H/H\) retains the local three-dimensional capture regime, but does not preserve the importance that disk self-gravity would have if included. Our simulations omit gas self-gravity, so the Toomre criterion is not a stability condition for them and does not set their numerical admissibility or stopping point. Their purpose is to constrain mass and angular-momentum delivery in this non-self-gravitating model. We report \(Q\) as a physical diagnostic for applying the model to real disks or extending it to include self-gravity.

For a self-gravitating gaseous disk the Toomre parameter is \(Q=c_s\kappa/(\pi G\Sigma)\), where \(\kappa\) is the radial epicyclic frequency of the axisymmetric background \citep{ref51}. The local, razor-thin axisymmetric criterion places instability below \(Q=1\); values of order unity also identify a regime in which gravitational transport can matter \citep{ref38}. For screening our stellar-dominated initial profiles, define \begin{equation}\label{eq:18}
\fitmath{\begin{gathered}
Q_{\rm K}=\frac{c_s\Omega_K}{\pi G\Sigma},\\[3pt]
 \Omega_K=\left(\frac{GM_\star}{R^3}\right)^{1/2}.
\end{gathered}}
\end{equation} This substitutes the stellar Keplerian frequency for the actual epicyclic frequency. It is an initial estimate, not a stability calculation for a finite-thickness, pressure-supported, evolving disk.

The surface-density maps in Table~\ref{tab:1} imply \begin{equation}\label{eq:19}
\fitmath{\frac{Q_{\rm K}'}{Q_{\rm K}}=
 \begin{cases}
 \Gamma^{-2},&\text{A: all radii},\\
 S^{-1},&\text{periodic sector}.
 \end{cases}}
\end{equation} Here the gas mass and sound speed are retained, and the initial column is mapped homologously. A denser reservoir thus gives faster delivery and clearing in our model while increasing the potential importance of the omitted self-gravity in a corresponding physical disk. For Scaling A the disk-to-central-mass ratio also rises as \(\Gamma^3\). For sector compression the local \(Q_{\rm K}\) estimate does not certify the stability of the periodically repeated domain. Applying the results to a disk where gravitational transport matters would require an assessment of that additional physics.

An initial \(Q_{\rm K}>1\) is not by itself evidence that self-gravity is negligible. Capture and dispersal can raise \(Q\), while local accumulation can lower it; total inventory reduction does not determine the minimum throughout the disk. The values reported below are therefore initial estimates, not evolved stability results.

Scaling B avoids the increased planet-to-star mass ratio, but brings periodic images closer and restricts the azimuthal modes. It cannot be pushed so far that the capture neighborhoods or the relevant flow structures no longer fit. Even before that geometrical limit, the shortened interval between encounters can substantially alter capture relative to passage (Section~\ref{sec:2.3}). Because this partition and its angular-momentum delivery are central to our purpose, we emphasize the full-annulus construction A. Its changed planetary separation and global geometry still enter the resulting flow; preferring A does not assume that its final capture fraction equals the unscaled system's.

Gap edges, pressure transport, meridional circulation, and vortices can respond to changed global geometry. Their response is part of the scaled simulation. The torque budget estimates the clearing time; the detailed gap profile also depends on pressure transport, as in \citet{ref3}. If a physical taper or Hill-width margin is not contracted, the clearing estimate must use the actual reservoir area.

These constraints guide the choice of scaled systems for sustained accretion and clearing simulations.

\section{Two applications of the scaling}\label{sec:3}

\subsection{Scaled PDS 70}\label{sec:3.1}

Take the PDS reference star and orbital radii, choose a globally isothermal cold disk, and set both initial planetary masses by \(R_H=H\). Apply all-radii compression as specified in Table~\ref{tab:2}. The reservoir exceeds the combined seed mass, allowing substantial growth and migration rather than only a perturbation about the seeds.

\begin{table*}[tp]
\centering
\caption{Scaled PDS 70 at $R_H=H$. Listed masses are nominal initial targets; realized bare seeds and post-capture states appear in Section~\ref{sec:4.6}. Gains are reference divided by scaled values; timestep and grid effects are separate.}\label{tab:2}
\small\setlength{\tabcolsep}{2pt}
\begin{tabular}{@{}lll@{}}
\toprule
\parbox[t]{0.4800\textwidth}{\raggedright Parameter\strut} & \parbox[t]{0.2400\textwidth}{\centering PDS reference\strut} & \parbox[t]{0.2400\textwidth}{\centering Scaled simulation\strut} \\[4pt]
\midrule
\parbox[t]{0.4800\textwidth}{\raggedright Compression \(\Gamma\)\strut} & \parbox[t]{0.2400\textwidth}{\centering \(1\)\strut} & \parbox[t]{0.2400\textwidth}{\centering \(1.75\)\strut} \\[4pt]
\parbox[t]{0.4800\textwidth}{\raggedright Central mass, \(M_\odot\)\strut} & \parbox[t]{0.2400\textwidth}{\centering \(0.76\)\strut} & \parbox[t]{0.2400\textwidth}{\centering \(0.142\)\strut} \\[4pt]
\parbox[t]{0.4800\textwidth}{\raggedright Inner / outer orbit, au\strut} & \parbox[t]{0.2400\textwidth}{\centering \(22.0/34.5\)\strut} & \parbox[t]{0.2400\textwidth}{\centering \(12.6/19.7\)\strut} \\[4pt]
\parbox[t]{0.4800\textwidth}{\raggedright Temperature, K\strut} & \parbox[t]{0.2400\textwidth}{\centering \(20\)\strut} & \parbox[t]{0.2400\textwidth}{\centering \(20\)\strut} \\[4pt]
\parbox[t]{0.4800\textwidth}{\raggedright Inner / outer seed, \(M_{\rm J}\)\strut} & \parbox[t]{0.2400\textwidth}{\centering \(0.264/0.518\)\strut} & \parbox[t]{0.2400\textwidth}{\centering \(0.264/0.518\)\strut} \\[4pt]
\parbox[t]{0.4800\textwidth}{\raggedright Inner / outer \(H=R_H\), au\strut} & \parbox[t]{0.2400\textwidth}{\centering \(1.06/2.07\)\strut} & \parbox[t]{0.2400\textwidth}{\centering \(1.06/2.07\)\strut} \\[4pt]
\parbox[t]{0.4800\textwidth}{\raggedright Gas mass, \(M_{\rm J}\)\strut} & \parbox[t]{0.2400\textwidth}{\centering \(2\)\strut} & \parbox[t]{0.2400\textwidth}{\centering \(2\)\strut} \\[4pt]
\parbox[t]{0.4800\textwidth}{\raggedright Nominal torque-budget half-clearing acceleration\strut} & \parbox[t]{0.2400\textwidth}{\centering \(1\)\strut} & \parbox[t]{0.2400\textwidth}{\centering \(3.06\)\strut} \\[4pt]
\parbox[t]{0.4800\textwidth}{\raggedright Equal-resolution grid saving\strut} & \parbox[t]{0.2400\textwidth}{\centering \(1\)\strut} & \parbox[t]{0.2400\textwidth}{\centering \(3.06\)\strut} \\[4pt]
\parbox[t]{0.4800\textwidth}{\raggedright Nominal work saving at equal timestep\strut} & \parbox[t]{0.2400\textwidth}{\centering \(1\)\strut} & \parbox[t]{0.2400\textwidth}{\centering \(9.38\)\strut} \\[4pt]
\parbox[t]{0.4800\textwidth}{\raggedright Conditional speedup if controlling velocity scales\strut} & \parbox[t]{0.2400\textwidth}{\centering \(1\)\strut} & \parbox[t]{0.2400\textwidth}{\centering \(16.4\)\strut} \\[4pt]
\bottomrule
\end{tabular}
\end{table*}

The nominal combined computational benefit is about ninefold \((\Gamma^4=9.38)\) at unchanged timestep and resolution. It consists of a 3.06-fold shorter physical half-clearing interval and a separate 3.06-fold reduction in cells. No exploratory coarsening is included in this gain. The comparison uses the matched reference in Table~\ref{tab:2}. If the atmospheric ceiling or orbital motion controls the timestep in both matched systems, the additional timestep benefit can approach 1.75. That gives the conditional 16.4-fold estimate, which depends on our numerical prescription (Section~\ref{sec:2.5}). We retain 9.38 as the nominal estimate and report any measured timestep contribution separately.

The selected reservoir is tapered between the planets without pre-imposed gaps or external replenishment. Stellar vertical stratification and pressure-supported rotation initialize the gas. Both planet potentials are present from the start; initial gap formation is a preparation phase whose transfers remain part of the subsequent state.

The selected radial domain is 7.89--26.6 au, with full azimuth and an upper wedge of 0.405 rad. A 1-au taper is retained as a practical profile choice rather than contracted homologously. The exact initialized profile enters the absolute clearing-time calculation. For the matched control, the homologous reference to this implemented profile has a 1.75-au taper. Table~\ref{tab:2} preserves the two-Jupiter-mass gas inventory before initial capture and gives nominal planetary masses; the realized seed and post-capture values are distinguished in Section~\ref{sec:4.6}. The exact reference normalization and native initial capture are recorded separately in each calculation.

The exploratory root mesh is \(36\times6\times112\), with \(12\times6\times16\) MeshBlocks and one added planet-following level. The finest local resolution is approximately \(4.1,\ 2.5,\ 3.0\) cells per \(H\) in the radial, polar, and azimuthal directions at the inner planet, and \(8.0,\ 3.1,\ 3.7\) at the outer planet. This is a deliberately coarse physical simulation, separate from the equal-resolution mesh comparison above.

Two sampled states identify velocity-dominated atmospheric timestep limits, one below the ceiling and one at it (Appendix~\ref{sec:A}). The matched benchmark measures the resulting mean-timestep ratio rather than inferring it from these individual states.

The seed targets include initial capture. The native discrete initialization realizes \(R_H/H=1\) to within \(3\times10^{-6}\), without counting that gas twice. The initial buffered patches are separated at conjunction. Even if the full reservoir were allocated to the planets at fixed orbits, a gap of native cells would remain between the patches. Migration remains an additional constraint. Appendix~\ref{sec:A} specifies the initializer and numerical geometry.

\subsection{Scaled Solar System}\label{sec:3.2}

Use a solar-mass reference star with Jupiter--Saturn orbital radii and the same illustrative cold temperature. The \(R_H=H\) rule now gives the seed masses in Table~\ref{tab:3}. The table compares four all-radii compression candidates for this 20-K isothermal disk.

\begin{table*}[tp]
\centering
\caption{All-radii scaled Solar-System candidates at $R_H=H$. Orbits are nominal inputs before initial capture; the realized starting orbits are reported in Section~\ref{sec:4.8}. Products of physical and grid gains assume unchanged timestep and cell-proportional cost. Hill clearances refer to seeds before buffers, growth, or migration.}\label{tab:3}
\small\setlength{\tabcolsep}{2pt}
\begin{tabular}{@{}llllll@{}}
\toprule
\parbox[t]{0.2600\textwidth}{\raggedright Parameter\strut} & \parbox[t]{0.1380\textwidth}{\centering Reference\strut} & \parbox[t]{0.1380\textwidth}{\centering \(\Gamma=3\)\strut} & \parbox[t]{0.1380\textwidth}{\centering \(\Gamma=4\)\strut} & \parbox[t]{0.1380\textwidth}{\centering \(\Gamma=5\)\strut} & \parbox[t]{0.1380\textwidth}{\centering \(\Gamma=6\)\strut} \\[4pt]
\midrule
\parbox[t]{0.2600\textwidth}{\raggedright Central mass, \(M_\odot\)\strut} & \parbox[t]{0.1380\textwidth}{\centering \(1\)\strut} & \parbox[t]{0.1380\textwidth}{\centering \(0.0370\)\strut} & \parbox[t]{0.1380\textwidth}{\centering \(0.0156\)\strut} & \parbox[t]{0.1380\textwidth}{\centering \(0.0080\)\strut} & \parbox[t]{0.1380\textwidth}{\centering \(0.00463\)\strut} \\[4pt]
\parbox[t]{0.2600\textwidth}{\raggedright Inner / outer orbit, au\strut} & \parbox[t]{0.1380\textwidth}{\centering \(5.20/9.58\)\strut} & \parbox[t]{0.1380\textwidth}{\centering \(1.73/3.19\)\strut} & \parbox[t]{0.1380\textwidth}{\centering \(1.30/2.40\)\strut} & \parbox[t]{0.1380\textwidth}{\centering \(1.04/1.92\)\strut} & \parbox[t]{0.1380\textwidth}{\centering \(0.867/1.60\)\strut} \\[4pt]
\parbox[t]{0.2600\textwidth}{\raggedright Temperature, K\strut} & \parbox[t]{0.1380\textwidth}{\centering \(20\)\strut} & \parbox[t]{0.1380\textwidth}{\centering \(20\)\strut} & \parbox[t]{0.1380\textwidth}{\centering \(20\)\strut} & \parbox[t]{0.1380\textwidth}{\centering \(20\)\strut} & \parbox[t]{0.1380\textwidth}{\centering \(20\)\strut} \\[4pt]
\parbox[t]{0.2600\textwidth}{\raggedright Inner / outer seed, \(M_{\rm J}\)\strut} & \parbox[t]{0.1380\textwidth}{\centering \(0.0264/0.0660\)\strut} & \parbox[t]{0.1380\textwidth}{\centering unchanged\strut} & \parbox[t]{0.1380\textwidth}{\centering unchanged\strut} & \parbox[t]{0.1380\textwidth}{\centering unchanged\strut} & \parbox[t]{0.1380\textwidth}{\centering unchanged\strut} \\[4pt]
\parbox[t]{0.2600\textwidth}{\raggedright Inner / outer \(H=R_H\), au\strut} & \parbox[t]{0.1380\textwidth}{\centering \(0.106/0.264\)\strut} & \parbox[t]{0.1380\textwidth}{\centering unchanged\strut} & \parbox[t]{0.1380\textwidth}{\centering unchanged\strut} & \parbox[t]{0.1380\textwidth}{\centering unchanged\strut} & \parbox[t]{0.1380\textwidth}{\centering unchanged\strut} \\[4pt]
\parbox[t]{0.2600\textwidth}{\raggedright Inner / outer \(H/a\)\strut} & \parbox[t]{0.1380\textwidth}{\centering \(0.0203/0.0276\)\strut} & \parbox[t]{0.1380\textwidth}{\centering \(0.0610/0.0828\)\strut} & \parbox[t]{0.1380\textwidth}{\centering \(0.0813/0.110\)\strut} & \parbox[t]{0.1380\textwidth}{\centering \(0.102/0.138\)\strut} & \parbox[t]{0.1380\textwidth}{\centering \(0.122/0.166\)\strut} \\[4pt]
\parbox[t]{0.2600\textwidth}{\raggedright Outer seed / central mass, \%\strut} & \parbox[t]{0.1380\textwidth}{\centering \(0.0063\)\strut} & \parbox[t]{0.1380\textwidth}{\centering \(0.170\)\strut} & \parbox[t]{0.1380\textwidth}{\centering \(0.403\)\strut} & \parbox[t]{0.1380\textwidth}{\centering \(0.788\)\strut} & \parbox[t]{0.1380\textwidth}{\centering \(1.36\)\strut} \\[4pt]
\parbox[t]{0.2600\textwidth}{\raggedright Initial Hill-surface clearance, au\strut} & \parbox[t]{0.1380\textwidth}{\centering \(4.01\)\strut} & \parbox[t]{0.1380\textwidth}{\centering \(1.09\)\strut} & \parbox[t]{0.1380\textwidth}{\centering \(0.725\)\strut} & \parbox[t]{0.1380\textwidth}{\centering \(0.506\)\strut} & \parbox[t]{0.1380\textwidth}{\centering \(0.360\)\strut} \\[4pt]
\parbox[t]{0.2600\textwidth}{\raggedright Nominal torque-budget half-clearing acceleration\strut} & \parbox[t]{0.1380\textwidth}{\centering \(1\)\strut} & \parbox[t]{0.1380\textwidth}{\centering \(9\)\strut} & \parbox[t]{0.1380\textwidth}{\centering \(16\)\strut} & \parbox[t]{0.1380\textwidth}{\centering \(25\)\strut} & \parbox[t]{0.1380\textwidth}{\centering \(36\)\strut} \\[4pt]
\parbox[t]{0.2600\textwidth}{\raggedright Equal-resolution grid saving\strut} & \parbox[t]{0.1380\textwidth}{\centering \(1\)\strut} & \parbox[t]{0.1380\textwidth}{\centering \(9\)\strut} & \parbox[t]{0.1380\textwidth}{\centering \(16\)\strut} & \parbox[t]{0.1380\textwidth}{\centering \(25\)\strut} & \parbox[t]{0.1380\textwidth}{\centering \(36\)\strut} \\[4pt]
\parbox[t]{0.2600\textwidth}{\raggedright Nominal work saving at equal timestep\strut} & \parbox[t]{0.1380\textwidth}{\centering \(1\)\strut} & \parbox[t]{0.1380\textwidth}{\centering \(81\)\strut} & \parbox[t]{0.1380\textwidth}{\centering \(256\)\strut} & \parbox[t]{0.1380\textwidth}{\centering \(625\)\strut} & \parbox[t]{0.1380\textwidth}{\centering \(1296\)\strut} \\[4pt]
\bottomrule
\end{tabular}
\end{table*}

The nominal combined computational gains range from 81-fold at \(\Gamma=3\) to 1296-fold at \(\Gamma=6\). In the latter case, a 36-fold shorter physical half-clearing interval combines with a separate 36-fold equal-resolution grid saving. Thus a thousandfold reduction in work is a potential benefit of aggressive compression, before any exploratory coarsening. The estimate combines the clearing-time scaling with an unchanged timestep and cell-proportional update cost, as derived in Section~\ref{sec:2.5}.

The useful upper compression is set by growth as well as initialization. Table~\ref{tab:3} gives conjunction clearances before numerical buffers or migration. The seed Hill spheres alone would touch near \(\Gamma=11.8\); the thickness, mass-hierarchy, and buffered-patch constraints of Section~\ref{sec:2.7} set the usable range during growth. For example, at unchanged scaled orbital radii, growth of the inner and outer planets to \(1\) and \(0.3\,M_{\rm J}\), respectively, would leave Hill-surface clearances of approximately 0.302 au at \(\Gamma=4\) and only 0.083 au at \(\Gamma=5\), before buffers. The inner planet would then be 6.1\% and 11.9\% of the central mass, respectively. This example quantifies the growth allowance needed when selecting the compression.

The reservoir mass limits which of these formal gains is physically useful. For a first \(2\,M_{\rm J}\) screening calculation, we adopt the PDS run's dimensionless radial domain and taper, now measured in units of the scaled outer orbit: \(0.4<r/a_c<1.35\), with taper centers at the planetary radii, width \(0.0507a_c\), and background amplitude \(10^{-4}\) of the reservoir amplitude. The upper wedge retains a height of approximately \(4.07H_c\) at the outer planet. We normalize the same isothermal vertical initializer over that spherical domain, counting both disk halves before capture. The first simulation uses this domain without an independent boundary extension.

This calculation gives initial minima \(Q_{{\rm K},\min}\simeq1.93\), 1.34, and 0.751 at \(\Gamma=2.5\), 3, and 4, respectively. At \(\Gamma=4\), the disk is 12.2\% of the central mass; at \(\Gamma=3\) it is 5.16\%. These values identify where self-gravity could become important in a physical counterpart (Section~\ref{sec:2.7}). The \(\Gamma=4\)--6 cost projections in Table~\ref{tab:3} should be interpreted within that same model.

We favor \(\Gamma=3\) as the first exploratory compromise: its nominal equal-timestep work saving is 81-fold, compared with 39-fold at \(\Gamma=2.5\). Its \(Q_{{\rm K},\min}\simeq1.34\) is retained as a physical diagnostic only. The exploratory root mesh is \(36\times6\times112\), with one planet-following refinement level and \(12\times6\times16\) MeshBlocks. The initialized native gas inventory is \(2.00\,M_{\rm J}\); adjusting the bare seeds for initially enclosed gas gives \(R_H/H\) within 0.25\% of unity. The initial target is 50\% inventory reduction, with a saved endpoint permitting a later decision on continuation; it is not a prescribed final planetary mass.

A closer pair can also be designed, but its separation is an additional physical choice. We do not mix that change into the four scaled columns above. Section~\ref{sec:4.8} reports the first Solar-System calculation with this exploratory mesh and the same delivery diagnostics, stopped near the half-inventory milestone. Its measured runtime establishes the practicality of this configuration; it is not a matched measurement of the nominal 81-fold scaling gain.

\section{Numerical model, diagnostics, and results}\label{sec:4}

\subsection{Exterior dynamics and perfect capture}\label{sec:4.1}

The simulations use Athena++'s three-dimensional hydrodynamics and block-based refinement \citep{ref7}. The gas is globally isothermal, inviscid, and non-self-gravitating. Both planetary potentials act on a shared exterior disk. Radial boundaries permit outflow but not imposed inflow; full azimuth is periodic. The upper wedge and reflecting midplane represent both disk halves by doubling gas inventories and transfers once.

Each sphere of influence is an excluded capture region with \(R_{\rm cap}=\min(R_B,R_H)\). Gas entering it is immediately retained. Its mass adds to the gravitating planet; its orbital and internal angular momenta are recorded separately. Inside it there is no evolved fluid, no resolved circumplanetary-disk (CPD) transport, and no re-emission. This perfect-capture limit defines the numerical question. The simulations test delivery to the sphere of influence, not accretion through a resolved CPD onto a planetary surface.

The thermal motivation is the rapid-cooling regime of gas-giant growth. \citet[\S{}4.2]{ref2} specifically cite \citet[\S{}2.1]{ref39} for cooling that accelerates once the envelope's self-gravity dominates. Dust opacity impedes radiative escape; grain growth and settling can lower that opacity and facilitate cooling. Detailed envelope calculations and grain-evolution models establish this mechanism \citep{ref41,ref42,ref43}. The nominal opacity adopted in those envelope models already incorporates grain growth, and their companion study explicitly discusses sedimentation and faster cooling at lower opacity \citep[\S{}\S{}2 and 5]{ref40}. We idealize efficient energy loss as immediate retention at the Hill surface, without a cooling delay or thermal feedback on the exterior flow. This is an imposed physical limit, not a radiative-transfer calculation. Captured mass and its angular momentum remain in the gravitating planet; we do not assume that its internal angular momentum disappears or that all retained gas has reached the planetary surface. Thus the calculation isolates delivery when cooling and subsequent retention do not throttle that delivery. We adopt this assumption for the present simulations and leave its thermal applicability to the companion theory paper; \(R_H=H\) specifies the initial delivery geometry, not a cooling criterion. Unresolved envelope self-gravity motivating this limit is distinct from the circumstellar gas self-gravity omitted from our exterior dynamics.

Moving exterior cut volumes and conservative merging handle the interface. Planet-following refinement is local; the domain is not refined uniformly. We use the second-order van Leer (VL2) time integrator \citep{ref52} with second-order spatial reconstruction, Harten--Lax--van Leer--Einfeldt (HLLE) Riemann-solver fluxes \citep{ref53,ref54}, and CFL number 0.3. Managed-patch aggregation does not retain native second-order reconstruction within every aggregated update, so its effect on capture must remain part of resolution assessment.

Both orbits remain circular, with radii updated from each planet's own exchanges with the gas. Direct planet--planet gravity is omitted. The calculation consequently does not test eccentricity excitation, resonant dynamics, or planet--planet stability. These omissions become particularly important when compression increases mass ratios or reduces separation.

\textbf{Scope.} These choices define what the simulations test: delivery to absorbing Hill spheres on circular orbits, without planet--planet gravity, gas self-gravity, or a resolved CPD. The meshes are exploratory: their coarsest local directions resolve the initial scale height with only a few cells, and conservation of the implemented exchanges does not establish convergence of the capture rate or of narrow inflow structures. The results below therefore measure bulk delivery and coupled evolution at the stated resolution; a resolution study has not been completed. Two further conventions apply throughout. The gas inventory between the moving planetary orbits (Section~\ref{sec:4.5}) changes through capture, radial displacement of gas, and migration of its bounding orbits, so its reduction is not physical clearing of the initial region. Comparison with inventories between the fixed initial orbits isolates the migration effect. Runtimes other than the matched benchmark of Section~\ref{sec:4.5} are practical costs, not speedup measurements.

\subsubsection{Linear torque benchmark}\label{sec:4.1.1}

We checked the gravitational torque calculation against the three-dimensional linear result of \citet{ref8} using a fixed, circular, non-accreting planet with mass ratio \(q=3.00\times10^{-5}\) and disk aspect ratio \(h=H/a=0.0811\). The globally isothermal disk has surface density proportional to \(R^{-1/2}\); the potential softening is \(0.100H\). The cylindrical root mesh contains \(72\times768\times72\) cells in \((R,\phi,z)\), with full azimuth and a vertical extent of \(\pm3H\). We compare this mesh with the same setup using two added local refinement levels around the planet.

\begin{figure*}[tp]\centering
\includegraphics[width=0.94\textwidth,height=0.50\textheight,keepaspectratio]{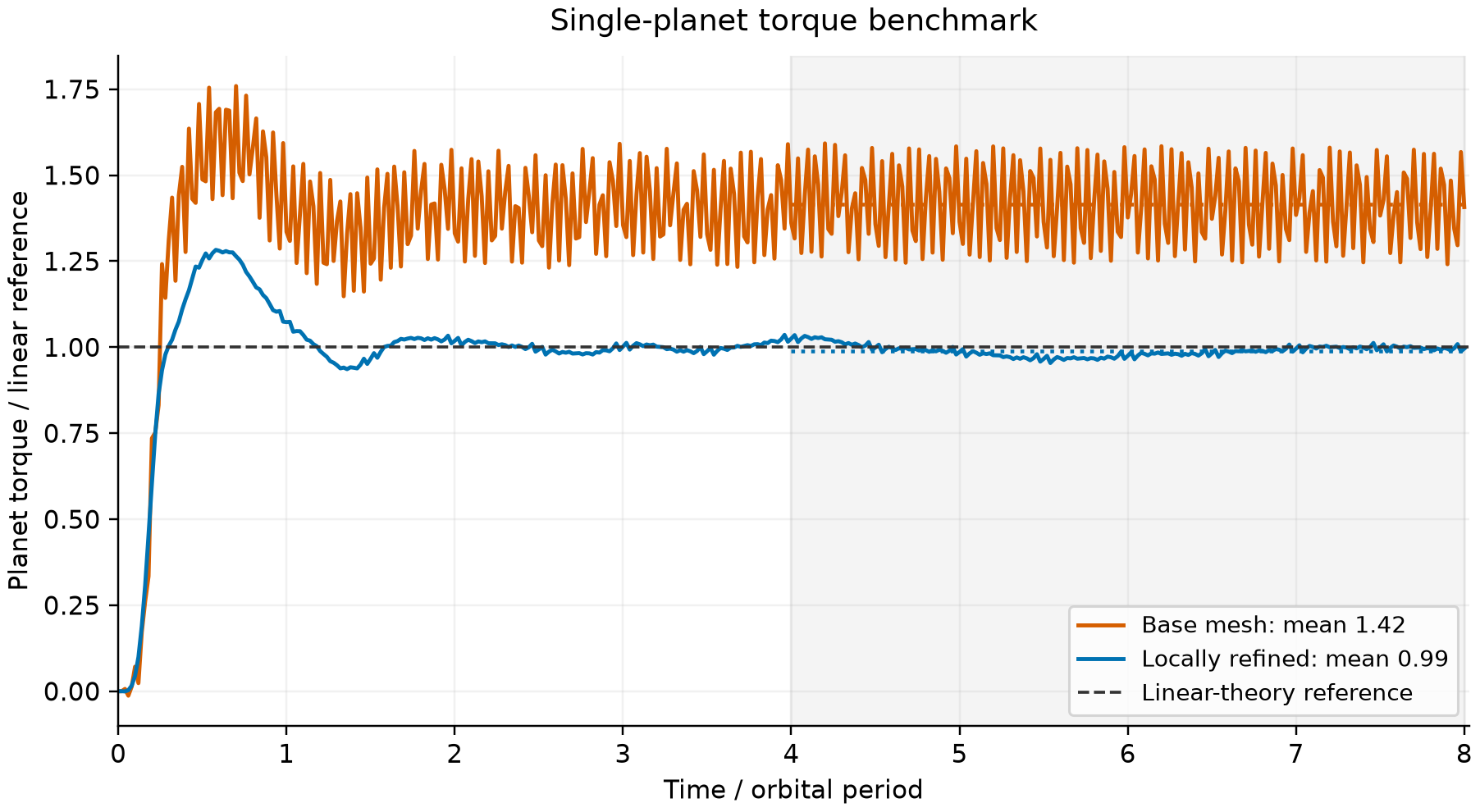}
\caption{Linear torque benchmark. The all-gas torque on a fixed planet, normalized by the Tanaka et al.~(2002) prediction, is shown for the base mesh and the locally refined calculation. The dashed line marks the analytical prediction; dotted colored lines mark the means over the shaded four-to-eight-orbit interval. Refinement changes the mean ratio from 1.42 to 0.990, recovering the linear torque to within 1\% without tuning a Hill taper.}\label{fig:1}
\end{figure*}

Figure~\ref{fig:1} shows the planet's gravitational torque, including all gas without a Hill-sphere exclusion or taper, divided by the Tanaka--Takeuchi--Ward reference torque \(\mathcal T_{\rm TTW}\). Let \(p_\Sigma=-1/2\) be the input surface-density exponent, \(\Sigma\propto R^{p_\Sigma}\), and let \(\overline\Sigma\) be the mean of the recorded local mass divided by its sampling area over the four-to-eight-orbit interval. This mean is measured separately for each mesh, not replaced by the initial surface density. The normalization used in the reduction is \begin{equation}\label{eq:20}
\fitmath{\mathcal T_{\rm TTW}=-(1.36-0.541p_\Sigma)
       \frac{q^2\overline\Sigma a^4\Omega^2}{h^2}.}
\end{equation} Writing \(\mathcal T_p\) for the measured planetary torque, averaging the 200 saved samples between four and eight orbits gives \(\langle\mathcal T_p\rangle/\mathcal T_{\rm TTW}=1.42\) on the base mesh and \(0.990\) with local refinement: the refined result agrees within 1\%. Both the measured mean torque and the reference are negative. This benchmark checks the underlying gravitational torque calculation in the linear regime and predates the excluded-sink capture treatment.

\subsection{Coupled mass and angular-momentum budgets}\label{sec:4.2}

Accretion-driven migration requires advected angular momentum and force torques to be accounted for separately. A related decomposition into gravity, accretion, pressure, and viscosity is used in the global two-dimensional calculations of \citet{ref49}. Our inviscid three-dimensional calculation additionally retains the captured gas's internal angular momentum in the planet.

For planet \(i\), let \(L_{o,i}\) be orbital angular momentum and \(J_i\) retained internal angular momentum. Over one accepted update, \(\Delta M_{{\rm cap},i}\geq0\) is the captured gas mass; \(\Delta L_{{\rm cap},i}\) and \(\Delta J_{{\rm cap},i}\) are its signed star-centered and planet-centered angular momenta, respectively. The pressure angular impulses \(I_{P,L,i}\) and \(I_{P,J,i}\) are the star-centered and internal transfers received by the planet at the absorbing interface. Positive values add angular momentum along the positive \(z\) axis to the corresponding ledger. In contrast, \(I_{g,i}\) is the signed gravitational angular impulse the planet gives the gas, so the planet receives \(-I_{g,i}\). The accepted exchanges are \begin{equation}\label{eq:21}
\fitmath{\begin{aligned}
 \Delta M_i&=\Delta M_{{\rm cap},i},\\
 \Delta J_i&=\Delta J_{{\rm cap},i}+I_{P,J,i},\\
 \Delta L_{o,i}&=-I_{g,i}+\Delta L_{{\rm cap},i}
                 -\Delta J_{{\rm cap},i}+I_{P,L,i}-I_{P,J,i},\\
 a_i&=\frac{(L_{o,i}/M_i)^2}{GM_\star}.
 \end{aligned}}
\end{equation} The pressure impulses are defined for the implemented absorbing interface. An exact spherical pressure traction has no planet-centered torque; a nonzero internal pressure impulse from the faceted representation is retained for conservation, not interpreted as physical CPD transport. The importance of sink angular-momentum handling is illustrated by \citet{ref50}, whose circumbinary simulations reduce sink-induced torque biases by leaving relative angular momentum in the gas. Our excluded-volume prescription is different: the captured internal angular momentum leaves the resolved gas but is retained explicitly in \(J_i\), rather than assigned to orbital motion.

Let \(L_{g,z}\) be the star-centered axial angular momentum of gas still on the grid, \(L_{\rm boundary,out}\) its cumulative signed boundary export, and \(L_0\) the initial total system angular momentum. Positive export adds the angular momentum carried out of the domain to the accounted total. The star-centered budget is \begin{equation}\label{eq:22}
\fitmath{\begin{aligned}
 L_{\rm accounted}&\equiv L_{g,z}+\sum_i(L_{o,i}+J_i)
                    +L_{\rm boundary,out},\\
 L_{\rm accounted}&=L_0+\Delta L_{\rm clip}+\epsilon_L.
 \end{aligned}}
\end{equation} Here \(\Delta L_{\rm clip}\) is the net contribution of velocity clipping to accepted evolution, and \(\epsilon_L\) represents the remaining numerical discrepancy. The component ceiling changes stored axial angular momentum when the azimuthal velocity is clipped. The density-floor operation itself leaves stored angular momentum unchanged, so it contributes no separate angular-momentum source in this equation. The runs do not separately record signed or absolute cumulative clipping: the measured raw residual is \(\Delta L_{\rm clip}+\epsilon_L\), not an attribution of that residual to clipping. Source attribution follows which planet exerted the force, not where gas ultimately ends up. Gas pushed past one planet can deliver angular momentum to the other before capture.

The quantity \begin{equation}\label{eq:23}
\fitmath{\begin{gathered}
\Delta\mathcal K_i=\Delta L_{{\rm cap},i}-\Delta J_{{\rm cap},i}
                 -\ell_i\Delta M_i,\\[3pt]
 \ell_i=L_{o,i}/M_i=\sqrt{GM_\star a_i},
\end{gathered}}
\end{equation} uses the pre-step orbital specific angular momentum \(\ell_i\) and measures the advective contribution to changing orbital specific angular momentum. Together with gravitational and pressure impulses, it connects delivery to migration and hence future feeding. Final masses and orbital evolution are not separate validation targets.

\subsection{Measuring the specific angular momentum of inflow}\label{sec:4.3}

The principal diagnostics are joint mass and angular-momentum delivery as functions of height and planet-relative azimuth, together with exterior transport and storage. Let \(\mathbf r,\mathbf v\) be the gas position and velocity in the star-centered inertial frame, and \(\mathbf r_i,\mathbf v_i\) those of planet \(i\) in the same frame. The subscript \(z\) selects the component normal to the disk midplane. The signed planet-centered specific angular momentum is \begin{equation}\label{eq:24}
\fitmath{j_{z,i}=[(\mathbf r-\mathbf r_i)\times(\mathbf v-\mathbf v_i)]_z.}
\end{equation} The associated circularization radius is \(r_c=j_{z,i}^2/(GM_i)\). These quantities characterize captured delivery, not a resolved CPD profile. Moments must retain signed cancellation, not only a characteristic circularization radius.

The central theoretical target is the incoming mass-flux-weighted specific angular momentum, not the angular momentum per unit mass of gas already stored inside a region. Let \(d\dot m_i\) denote the positive inward mass flux through the moving capture surface, including its motion and expansion. With \(\ell_z=(\mathbf r\times\mathbf v)_z\), define \begin{equation}\label{eq:25}
\fitmath{\begin{aligned}
 \dot M_i&=\int d\dot m_i,\\
 \bar j_{{\rm in},i}&=\frac{\int j_{z,i}\,d\dot m_i}{\dot M_i},\\
 \bar\ell_{{\rm in},i}&=\frac{\int\ell_z\,d\dot m_i}{\dot M_i}.
 \end{aligned}}
\end{equation} The first angular momentum is planet-centered; the second is star-centered. Both use the same captured contributions and the same inertial frame. Height and azimuth bins reveal which streams supply each average. For the internal delivery, a useful dimensionless measurement is \begin{equation}\label{eq:26}
\fitmath{\widehat j_i=\frac{\bar j_{{\rm in},i}}{\Omega_iR_{H,i}^2}.}
\end{equation} The denominator is preserved by the scaling at corresponding initial states. The measured value of \(\widehat j_i\) directly tests the theory's prediction of angular-momentum delivery. Vertical motion affects which gas crosses the surface and its flux weight, even though the axial angular momentum itself depends on horizontal position and velocity.

The height dependence has a dynamical interpretation through stellar tidal torques. In the late gap-fed ballistic calculation of \citet{ref6}, trajectories launched above and below the midplane begin with the same axial specific angular momentum as their matched midplane trajectories, but acquire lower angular momentum on average before reaching the adopted interaction surface. In the local Hill approximation, the stellar contribution obeys \(d j_z/dt=-3\Omega_i^2xy\), where \(x,y\) are dimensional planet-centered coordinates in the rotating orbital plane, with \(x\) directed toward the star and \(y\) completing a right-handed frame with \(z\). Planetary gravity is central and supplies no planet-centered torque. Height nevertheless changes the horizontal trajectory, because planetary gravity depends on the full distance from the planet, and so changes the accumulated stellar torque. The reduction is therefore a dynamical result for that ensemble, not simply a shorter lever arm at entry. Ballistic trajectories may approximate angular-momentum delivery, but neglect pressure torques, which our hydrodynamic simulations retain. The simulations measure height-dependent delivery at the Hill capture surface, without resolving the subsequent interior evolution.

Passing gas requires exterior flux and encounter-region storage measurements. It cannot be reconstructed from a capture journal alone. Likewise, gas remaining on the grid may cease contributing to growth while retaining angular momentum and continuing to torque the planets. The analysis must distinguish capture, passage, storage, and domain escape without treating each surface crossing as permanent capture or permanent loss.

The scaled simulations record compact budget totals at every accepted step rather than a full per-event history. Joint capture-bin sums are accumulated at every step and written roughly every 20 steps, so no contribution is discarded; selected exterior snapshots and rolling restart states complete the record. Bins should be commensurate with the exploratory mesh.

\subsection{What the delivery theory must predict}\label{sec:4.4}

The theory represents the evolving shared reservoir and the reciprocal torque exchanges, then adds a meridional delivery description to distinguish gas captured by each planet from gas that passes it. The proposed simple starting point reconstructs height-dependent flow on a prescribed vertical column using an algebraic response, rather than solving another full three-dimensional fluid problem. Its task is to predict both how much gas arrives and the specific angular momentum carried by that gas. The numerical simulation constrains the reconstruction through \(\dot M_i\), \(\bar j_{{\rm in},i}\), and \(\bar\ell_{{\rm in},i}\), not through mass growth alone.

For example, two reconstructions can deliver the same total mass while assigning different proportions to high-altitude and near-midplane streams. If those streams carry different angular momenta, the growth rates agree but the predicted CPD supply and orbital response do not. Flux-weighted specific angular momentum exposes this difference. The retained planet-centered component also supplies a circularization scale, although the distribution and signed cancellation must be retained rather than replaced by one mean radius.

The same captured contribution removes mass and angular momentum from the gas and supplies the planet's orbital and internal ledgers. Passing material remains in the exterior budget and can continue interacting with either planet. Growth, orbital response, and changing access to the reservoir are then evolved together. The target is not merely a fitted total accretion rate: it is a useful prediction of this coupled partition. In the adopted orbital/internal allocation, the specific angular momentum assigned to the orbit by capture is \begin{equation}\label{eq:27}
\fitmath{\begin{gathered}
\ell_{{\rm cap},o,i}=\bar\ell_{{\rm in},i}-\bar j_{{\rm in},i},\\[3pt]
 \left(\frac{\dot a_i}{a_i}\right)_{\rm cap}
 =2\frac{\dot M_i}{M_i}
 \left(\frac{\ell_{{\rm cap},o,i}}{\ell_i}-1\right).
\end{gathered}}
\end{equation} Gravitational and pressure torques contribute separately as in Section~\ref{sec:4.2}. Excess orbital angular momentum per unit captured mass moves the circular orbit outward; a deficit moves it inward. This changes which gas the planet can subsequently reach. Constraining inflow angular momentum therefore constrains future growth and the final mass partition, not merely CPD size.

The response equations, coefficient choices, prior reduced-model comparisons, and possible extensions belong to the companion theory paper \citep{ref28}. Here we specify the measurements and comparisons that can support or challenge that description.

\subsection{Matched benchmark and extended inventory evolution}\label{sec:4.5}

Define the exterior inventory between the instantaneous circular orbits by \begin{equation}\label{eq:28}
\fitmath{\begin{gathered}
M_{\rm between}(t)=
 \int_{\substack{a_b(t)<R<a_c(t)\\{\rm exterior}}}\rho\,dV,
 \\[3pt]
 t_{1/2,{\rm inv}}=\inf\left\{t:
 M_{\rm between}(t)\leq\tfrac12\,M_{\rm between}(0^+)\right\}.
\end{gathered}}
\end{equation} The reference \(0^+\) follows initial capture and is fixed thereafter. Here initial capture means gas already present inside the planets' initial Hill spheres: it is removed from the evolved exterior and assigned to the planet--CPD ledgers during initialization. It is not subsequent inflow across the capture surfaces. This gas is retained in the mass and angular-momentum budgets, but is separated from the later accretion history. The volume integral includes both represented disk halves and excludes the capture interiors. In radial screening models the analogous integral is a sum over active annuli.

In the inventory figures, ``grid gas interior to the planets'' denotes gas between the two orbits, and ``grid gas exterior to the planets'' denotes gas on either radial side of that interval. These pair-wide labels differ from the starward and outward sides of an individual orbit used for the torque decomposition in Section~\ref{sec:5.2}. Gas exterior to a capture volume simply means gas outside the absorbing sphere.

Reduction of this inventory provides a common comparison horizon, but it is not physical clearing of the original interplanetary region: capture, radial displacement, and motion of the bounding planetary orbits all change the inventory, and gas removed from it can remain on the grid. We therefore use ``clearing'' for the physical process and torque-based estimates, and ``inventory reduction'' for this moving-orbit diagnostic. Inventories between the fixed initial orbits, together with total exterior gas, captured mass, boundary escape, and radial profiles, separate physical removal from orbital-boundary reclassification (Sections~\ref{sec:4.6}--\ref{sec:4.8}).

We distinguish a short matched computational benchmark from the extended scaled demonstration. For compression factor \(\Gamma\), define \begin{equation}\label{eq:29}
\fitmath{D_\Gamma(t)=1-\frac{M_{{\rm between},\Gamma}(t)}
                        {M_{{\rm between},\Gamma}(0^+)}.}
\end{equation} Each run uses its own post-initial-capture inventory as the fixed reference. The benchmark endpoint \(t_{1,\Gamma}\) is the first scheduled diagnostic time at which the conservative inventory bounds certify \(D_\Gamma\geq0.01\); it can therefore slightly follow the exact first crossing, and the realized reductions slightly exceed 1\%. The comparison reaches this target in 66.2 minutes unscaled and 7.12 minutes with Scaling A at \(\Gamma=1.75\), a 9.31-fold saving. Both calculations started afresh and ran sequentially on the same computer with eight Message Passing Interface (MPI) ranks, one thread per rank, and no competing heavy simulation. They use the identical executable, controller, capture algorithm, and fractional-volume diagnostic; Appendix~\ref{sec:A.5} identifies the hardware and software build. Temperature, physical bare seed masses, represented gas mass, and the dimensionless CFL and velocity-ceiling prescriptions are retained under the Scaling A parameter mapping.

Table~\ref{tab:4} reports the measured costs. Timings run from controller startup to the check that confirms the inventory target, including segment startup, coupling trials, checkpoint I/O, and endpoint checks, but not preparation of the inputs and executable before launch. Both runs use the same 20-step segmentation and output settings. The wall-clock durations of the 134 and 13 MPI-launched segments sum to 65.4 and 7.05 minutes; they include startup and I/O and are not isolated hydrodynamic-update timings.

\begin{table*}[tp]
\centering
\caption{Eight-rank benchmark through the 1\% inventory target. Both runs use the same executable and 20-step stopping cadence. Bounds describe fractional-volume integration and are rounded outward; stopping uses unrounded values. Elapsed time includes startup, checkpoint I/O, and endpoint checks. Actual target overshoots are retained. $^*$Gas accreted excludes gas already inside the Hill spheres at initialization, which is assigned to the respective planets before evolution begins. Accreted gas includes material retained in the CPD.}\label{tab:4}
\small\setlength{\tabcolsep}{2pt}
\begin{tabular}{@{}lll@{}}
\toprule
\parbox[t]{0.4800\textwidth}{\raggedright Quantity\strut} & \parbox[t]{0.2400\textwidth}{\centering Unscaled reference\strut} & \parbox[t]{0.2400\textwidth}{\centering Scaling A, \(\Gamma=1.75\)\strut} \\[4pt]
\midrule
\parbox[t]{0.4800\textwidth}{\raggedright MPI ranks\strut} & \parbox[t]{0.2400\textwidth}{\centering 8\strut} & \parbox[t]{0.2400\textwidth}{\centering 8\strut} \\[4pt]
\parbox[t]{0.4800\textwidth}{\raggedright Root mesh \((r,\theta,\phi)\)\strut} & \parbox[t]{0.2400\textwidth}{\centering \(60\times6\times192\)\strut} & \parbox[t]{0.2400\textwidth}{\centering \(36\times6\times112\)\strut} \\[4pt]
\parbox[t]{0.4800\textwidth}{\raggedright Root cells\strut} & \parbox[t]{0.2400\textwidth}{\centering 69,120\strut} & \parbox[t]{0.2400\textwidth}{\centering 24,192\strut} \\[4pt]
\parbox[t]{0.4800\textwidth}{\raggedright Mean MeshBlocks, step-weighted (1,152 cells each)\strut} & \parbox[t]{0.2400\textwidth}{\centering 164.6\strut} & \parbox[t]{0.2400\textwidth}{\centering 91.5\strut} \\[4pt]
\parbox[t]{0.4800\textwidth}{\raggedright Coupling trials per accepted step\strut} & \parbox[t]{0.2400\textwidth}{\centering 5.0\strut} & \parbox[t]{0.2400\textwidth}{\centering 5.8\strut} \\[4pt]
\parbox[t]{0.4800\textwidth}{\raggedright Accepted steps\strut} & \parbox[t]{0.2400\textwidth}{\centering 2,680\strut} & \parbox[t]{0.2400\textwidth}{\centering 260\strut} \\[4pt]
\parbox[t]{0.4800\textwidth}{\raggedright Physical duration (yr)\strut} & \parbox[t]{0.2400\textwidth}{\centering 52.4\strut} & \parbox[t]{0.2400\textwidth}{\centering 16.2\strut} \\[4pt]
\parbox[t]{0.4800\textwidth}{\raggedright Mean accepted timestep (yr)\strut} & \parbox[t]{0.2400\textwidth}{\centering 0.0195\strut} & \parbox[t]{0.2400\textwidth}{\centering 0.0624\strut} \\[4pt]
\parbox[t]{0.4800\textwidth}{\raggedright Total elapsed time (min)\strut} & \parbox[t]{0.2400\textwidth}{\centering 66.2\strut} & \parbox[t]{0.2400\textwidth}{\centering 7.12\strut} \\[4pt]
\parbox[t]{0.4800\textwidth}{\raggedright Mean elapsed cost per step (s)\strut} & \parbox[t]{0.2400\textwidth}{\centering 1.48\strut} & \parbox[t]{0.2400\textwidth}{\centering 1.64\strut} \\[4pt]
\parbox[t]{0.4800\textwidth}{\raggedright Initial interplanetary inventory, \(0^+\) (\(M_{\rm J}\))\strut} & \parbox[t]{0.2400\textwidth}{\centering 1.80\strut} & \parbox[t]{0.2400\textwidth}{\centering 1.78\strut} \\[4pt]
\parbox[t]{0.4800\textwidth}{\raggedright Inventory reduction (\%)\strut} & \parbox[t]{0.2400\textwidth}{\centering 1.01\strut} & \parbox[t]{0.2400\textwidth}{\centering 1.06\strut} \\[4pt]
\parbox[t]{0.4800\textwidth}{\raggedright Geometric reduction bounds (\%; rounded outward)\strut} & \parbox[t]{0.2400\textwidth}{\centering 1.00--1.02\strut} & \parbox[t]{0.2400\textwidth}{\centering 1.03--1.08\strut} \\[4pt]
\parbox[t]{0.4800\textwidth}{\raggedright Interplanetary inventory decrease (\(M_{\rm J}\))\strut} & \parbox[t]{0.2400\textwidth}{\centering 0.0182\strut} & \parbox[t]{0.2400\textwidth}{\centering 0.0188\strut} \\[4pt]
\parbox[t]{0.4800\textwidth}{\raggedright Gas accreted\(^*\), inner planet (\(M_{\rm J}\))\strut} & \parbox[t]{0.2400\textwidth}{\centering 0.00398\strut} & \parbox[t]{0.2400\textwidth}{\centering 0.00318\strut} \\[4pt]
\parbox[t]{0.4800\textwidth}{\raggedright Gas accreted\(^*\), outer planet (\(M_{\rm J}\))\strut} & \parbox[t]{0.2400\textwidth}{\centering 0.00861\strut} & \parbox[t]{0.2400\textwidth}{\centering 0.00579\strut} \\[4pt]
\parbox[t]{0.4800\textwidth}{\raggedright Total gas accreted\(^*\) (\(M_{\rm J}\))\strut} & \parbox[t]{0.2400\textwidth}{\centering 0.0126\strut} & \parbox[t]{0.2400\textwidth}{\centering 0.00897\strut} \\[4pt]
\parbox[t]{0.4800\textwidth}{\raggedright Gas accreted\(^*\) / inventory decrease (\%)\strut} & \parbox[t]{0.2400\textwidth}{\centering 69.2\strut} & \parbox[t]{0.2400\textwidth}{\centering 47.6\strut} \\[4pt]
\parbox[t]{0.4800\textwidth}{\raggedright Inner share of gas accreted\(^*\) (\%)\strut} & \parbox[t]{0.2400\textwidth}{\centering 31.6\strut} & \parbox[t]{0.2400\textwidth}{\centering 35.4\strut} \\[4pt]
\bottomrule
\end{tabular}
\end{table*}

Scaling A accretes 29\% less gas by this early inventory milestone, while the inner planet receives a larger share of the accreted mass: 35.4\% versus 31.6\% unscaled. The ratio of gas accreted to interplanetary inventory decrease is 47.6\% scaled and 69.2\% unscaled. The denominator is a net decrease between the moving orbits, not a tagged population of gas that has crossed a planetary orbit, so its unaccreted remainder cannot all be assigned to passage: migration also changes the inventory.

One additional local refinement level is used in each case. MeshBlock rounding gives slightly different local physical widths: the reference cells are 5.0\%, 3.8\%, and 2.1\% wider in the radial, polar, and azimuthal directions, respectively. At the initial inner orbit the finest widths \((\Delta r,r\Delta\theta,r\Delta\phi)\) are \((0.273,0.440,0.360)\) au unscaled and \((0.260,0.424,0.353)\) au scaled; at the outer orbit they are \((0.273,0.690,0.565)\) and \((0.260,0.665,0.553)\) au. The corresponding cells per \(H\) are \((3.86,2.40,2.93)\) and \((4.06,2.49,2.99)\) at the inner planet, and \((7.59,3.00,3.67)\) and \((7.97,3.12,3.75)\) at the outer planet. These are also the nominal cells per initial \(R_H\) for the \(R_H=H\) design, before discrete initial capture. Thus the benchmark approximately matches local resolution; its gain does not come from coarsening the scaled grid. Table~\ref{tab:12} gives the reference and scaled physical geometries.

The stopping diagnostic uses fractional overlap of native exterior cell volumes with the instantaneous orbital interval, with 64 polar strips to bound partially intersected volumes. Its post-initial-capture references are \(1.80\,M_{\rm J}\) unscaled and \(1.78\,M_{\rm J}\) scaled, as listed in Table~\ref{tab:4}. The latter is not the \(1.80\,M_{\rm J}\) cell-center inventory of the separate extended scaled run in Section~\ref{sec:4.6}; the runs use their own initialized states and inventory estimators. Native cell mass is distributed over its faceted exterior volume without applying the cut fraction a second time. These bounds concern geometric integration of the saved cell states, not unresolved density structure. Every 20 accepted steps the upper bound on remaining inventory, divided by the lower initial-inventory bound, is compared with 0.99. Both reported endpoints satisfy that conservative test. The last nonqualifying and first qualifying checks occur at cycles 2,660--2,680 unscaled and 240--260 scaled. Table~\ref{tab:4} reports the resulting small overshoots rather than interpolating to an exact 1\% crossing.

Using these reached endpoints, the duration, mean-step-cost, and mean-timestep factors in Section~\ref{sec:2.5} are respectively 3.23, 0.903, and 3.19: \begin{equation}\label{eq:30}
\fitmath{\mathcal S_1=3.23\times0.903\times3.19\simeq9.31.}
\end{equation} The reported speedup is evaluated before rounding these factors. The shorter physical interval and larger mean timestep together reduce the accepted step count by a factor 10.3. Despite the smaller domain, the scaled mean cost per accepted step is about 11\% higher, not lower.

In terms of the effective compressions of Section~\ref{sec:2.5}, the measured inventory-duration gain gives \(\Gamma_{\rm dur}=1.80\), close to \(\Gamma=1.75\). This endpoint includes capture and migration as well as gas clearing, so the agreement does not separately verify the torque-only clearing law. The grid saving is only partly realized. Weighting the MeshBlock count at the end of each 20-step segment by its steps, the meshes average 164.6 blocks unscaled and 91.5 scaled (Table~\ref{tab:4}), so \(\Gamma_{\rm grid}=1.34\). The root meshes differ by a factor 2.86, close to \(\Gamma^2\), but the refined regions around the planets keep their physical size because \(R_H\) is preserved. The scaled mesh starts at 70 blocks and ranges over 77--98 during the run; the unscaled mesh starts at 144 and ranges over 130--186. The endpoint counts, 98 and 130, are therefore not representative of the run. The remaining factor, \(\eta_{\rm cost}\simeq0.50\), shows that the scaled run costs about twice as much per cell-step: coupling trials, capture-patch work, and segment overhead do not shrink with the domain. Evaluated before rounding, \(\mathcal S_1=\Gamma_{\rm dur}^2\,\Gamma_{\rm grid}^2\,\eta_{\rm cost}\,\chi_1\simeq1.80^2\times1.34^2\times0.50\times3.19\simeq9.31\). Consequently, the agreement of the total saving with \(\Gamma^4=9.38\) does not establish the separate cell-proportional-cost and equal-timestep assumptions used in that nominal estimate. Nor does the measured timestep ratio by itself identify which cells or velocity components controlled the CFL condition. That ratio, \(\chi_1=3.19\), exceeds \(\Gamma\), the largest factor the velocity ceiling can supply to corresponding cells of equal width (Section~\ref{sec:2.5}); it therefore reflects differences in the realized flow and limiting-cell history as well as the adopted ceiling prescription.

This gain is measured early in the evolution, at a common inventory target. The scaled and unscaled systems need not share the same mass partition or orbital evolution, and the target includes migration of the bounding orbits as well as gas removal. The extended scaled PDS calculation of Section~\ref{sec:4.6} is not used as the timing denominator because competing workloads were not controlled. It demonstrates substantial evolution, reaching the 50\% inventory-reduction target, without imposing that long endpoint on the uncompressed control.

\subsection{Scaled PDS simulation}\label{sec:4.6}

\begin{figure*}[tp]\centering
\includegraphics[width=0.94\textwidth,height=0.50\textheight,keepaspectratio]{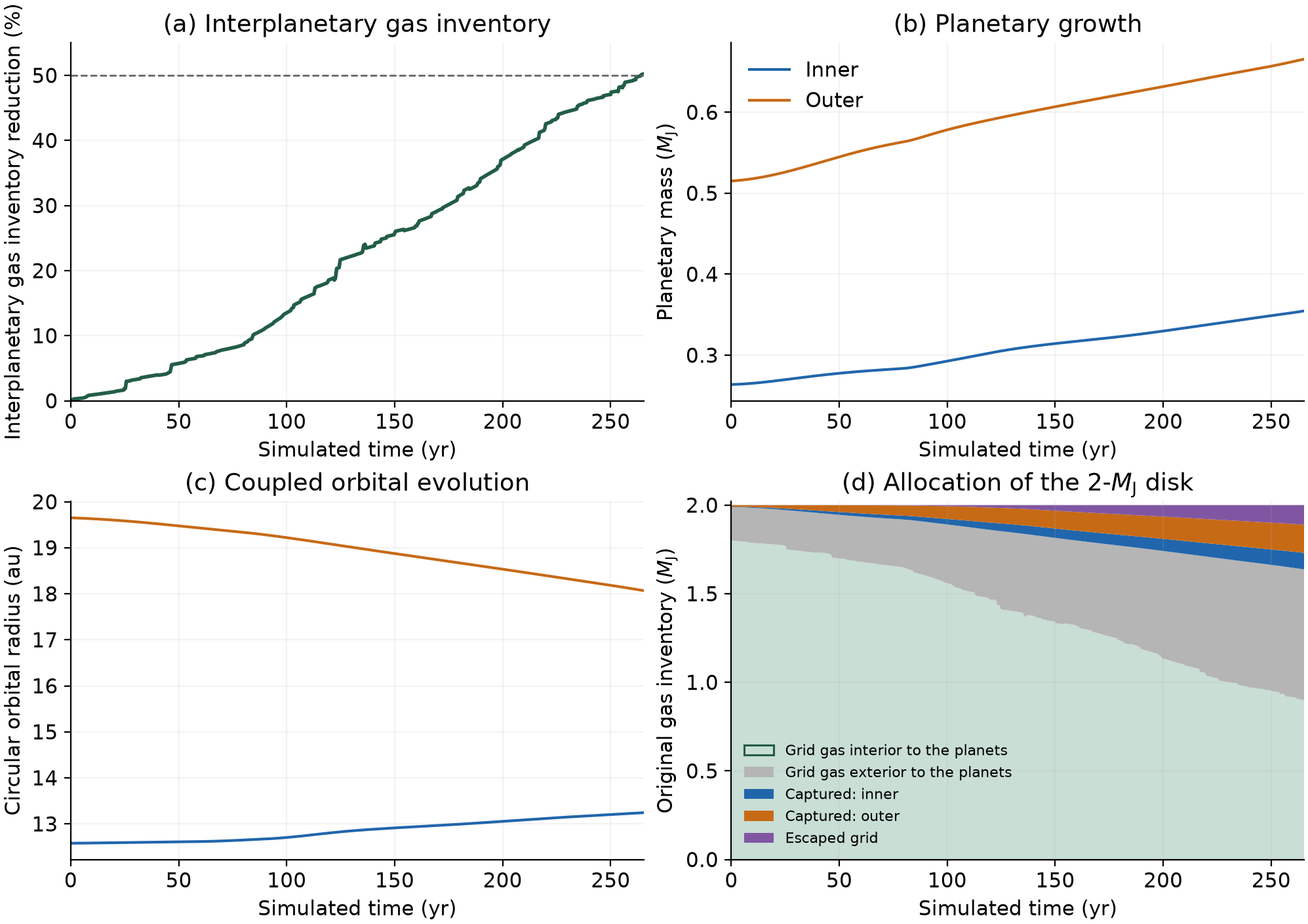}
\caption{Scaled PDS evolution. (a) Reduction of the gas inventory between the moving planetary orbits, relative to its post-initial-capture value of \(1.80\,M_{\rm J}\); the dashed line marks the 50\% stopping target. (b) Planetary masses, including retained CPD gas. (c) Circular orbital radii. (d) Partition of the original \(2\,M_{\rm J}\) disk. Green (grid gas interior to the planets) lies between the moving orbits; gray (grid gas exterior to the planets) lies inside the inner orbit or beyond the outer orbit and is not boundary loss. Capture inventories include gas already inside the initial Hill spheres at zero time. Small steps in (a) occur as the orbits cross cell centers.}\label{fig:2}
\end{figure*}

The coarse \(\Gamma=1.75\) calculation with a \(2\,M_{\rm J}\) reservoir halves the gas inventory between the moving planetary orbits in about 265 yr of simulated time (Figure~\ref{fig:2}). The inventory first falls by 50\% near 264 yr; the final saved endpoint, at 265 yr, has a 50.2\% reduction. Over this interval the planets grow from \(0.264\) and \(0.515\,M_{\rm J}\) to \(0.355\) and \(0.666\,M_{\rm J}\), and their orbits converge: the inner radius increases from 12.6 to 13.2 au and the outer radius decreases from 19.7 to 18.1 au. The inner planet receives 37.6\% of the mass captured after initialization. The calculation took 17.6 hours of wall-clock time, about 239 rank-hours on a root mesh of 24,192 cells (Appendix~\ref{sec:A.5}). Compression, the smaller domain, and the chosen coarse resolution all contribute to this cost (Section~\ref{sec:2.5}); the matched speedup is the separate measurement of Section~\ref{sec:4.5}.

\textbf{Which mass is being halved?} The original disk contains \(2.00\,M_{\rm J}\). After initialization, \(1.80\,M_{\rm J}\) lies between the planetary orbits, \(0.186\,M_{\rm J}\) lies outside them, and \(0.0113\,M_{\rm J}\) that lay inside the initial Hill spheres has been assigned to the capture ledgers. The reference for inventory reduction is therefore \(M_{\rm between}(0^+)=1.80\,M_{\rm J}\), not the whole disk. At the endpoint \(0.897\,M_{\rm J}\) remains between the orbits; the decrease of \(0.905\,M_{\rm J}\) meets the 50\% target. Halving this inventory does not mean that half the disk has been accreted or has left the computational domain.

\textbf{Mass partition.} Table~\ref{tab:5} partitions the original disk, excluding the seed planets and counting both disk halves once. Capture after initialization supplies \(0.241\,M_{\rm J}\), while grid gas outside the planetary orbits increases from \(0.186\) to \(0.740\,M_{\rm J}\), a net gain of \(0.554\,M_{\rm J}\), or about 2.3 times the accreted mass. This comparison subtracts the initial exterior inventory and excludes escape through the grid boundaries, which Table~\ref{tab:5} lists separately; escape is a domain-budget entry, not a separate planetary-feeding outcome.

\begin{table}[tp]
\centering
\caption{Partition of the original two-Jupiter-mass gas disk. Initial capture assigns gas already inside the initial Hill spheres to the planets; it is not later inflow. The saved endpoint is cycle 11,912 (265 yr). Seeds are excluded. Both disk halves are included once. Rounding can affect the displayed sum.}\label{tab:5}
\small\setlength{\tabcolsep}{2pt}
\begin{tabular}{@{}lll@{}}
\toprule
\parbox[t]{0.4800\columnwidth}{\raggedright Gas inventory, \(M_{\rm J}\)\strut} & \parbox[t]{0.2400\columnwidth}{\centering After initial capture\strut} & \parbox[t]{0.2400\columnwidth}{\centering Final diagnostic\strut} \\[4pt]
\midrule
\parbox[t]{0.4800\columnwidth}{\raggedright Grid gas interior to the planets\strut} & \parbox[t]{0.2400\columnwidth}{\centering 1.80\strut} & \parbox[t]{0.2400\columnwidth}{\centering 0.897\strut} \\[4pt]
\parbox[t]{0.4800\columnwidth}{\raggedright Grid gas exterior to the planets\strut} & \parbox[t]{0.2400\columnwidth}{\centering 0.186\strut} & \parbox[t]{0.2400\columnwidth}{\centering 0.740\strut} \\[4pt]
\parbox[t]{0.4800\columnwidth}{\raggedright Captured by inner planet\strut} & \parbox[t]{0.2400\columnwidth}{\centering 0.0022\strut} & \parbox[t]{0.2400\columnwidth}{\centering 0.0931\strut} \\[4pt]
\parbox[t]{0.4800\columnwidth}{\raggedright Captured by outer planet\strut} & \parbox[t]{0.2400\columnwidth}{\centering 0.0090\strut} & \parbox[t]{0.2400\columnwidth}{\centering 0.160\strut} \\[4pt]
\parbox[t]{0.4800\columnwidth}{\raggedright Net escape through grid boundaries\strut} & \parbox[t]{0.2400\columnwidth}{\centering 0\strut} & \parbox[t]{0.2400\columnwidth}{\centering 0.110\strut} \\[4pt]
\parbox[t]{0.4800\columnwidth}{\raggedright Total original gas\strut} & \parbox[t]{0.2400\columnwidth}{\centering 2.00\strut} & \parbox[t]{0.2400\columnwidth}{\centering 2.00\strut} \\[4pt]
\bottomrule
\end{tabular}
\end{table}

\textbf{Moving and fixed intervals.} The \(0.554\,M_{\rm J}\) exterior increase has two contributions. Gas can move beyond the initial orbits, but the orbits also move: the outer orbit contracts and the inner orbit expands, so gas in the strips they sweep is reclassified as exterior without itself moving radially. Let \(E_0(t)\) be the exterior gas on the grid classified with the fixed post-initialization radii \(a_b(0^+)\) and \(a_c(0^+)\), and let \(E(t)\) be the same gas classified with the instantaneous radii, as in Figure~\ref{fig:2}. Then \begin{equation}\label{eq:31}
\fitmath{\begin{aligned}
 E(t)-E_0(0^+)&=[E_0(t)-E_0(0^+)]\\
              &\quad+[E(t)-E_0(t)].
 \end{aligned}}
\end{equation} The first term is net accumulation outside the original orbits; the second is reclassification by migration.

\begin{figure*}[tp]\centering
\includegraphics[width=0.94\textwidth,height=0.50\textheight,keepaspectratio]{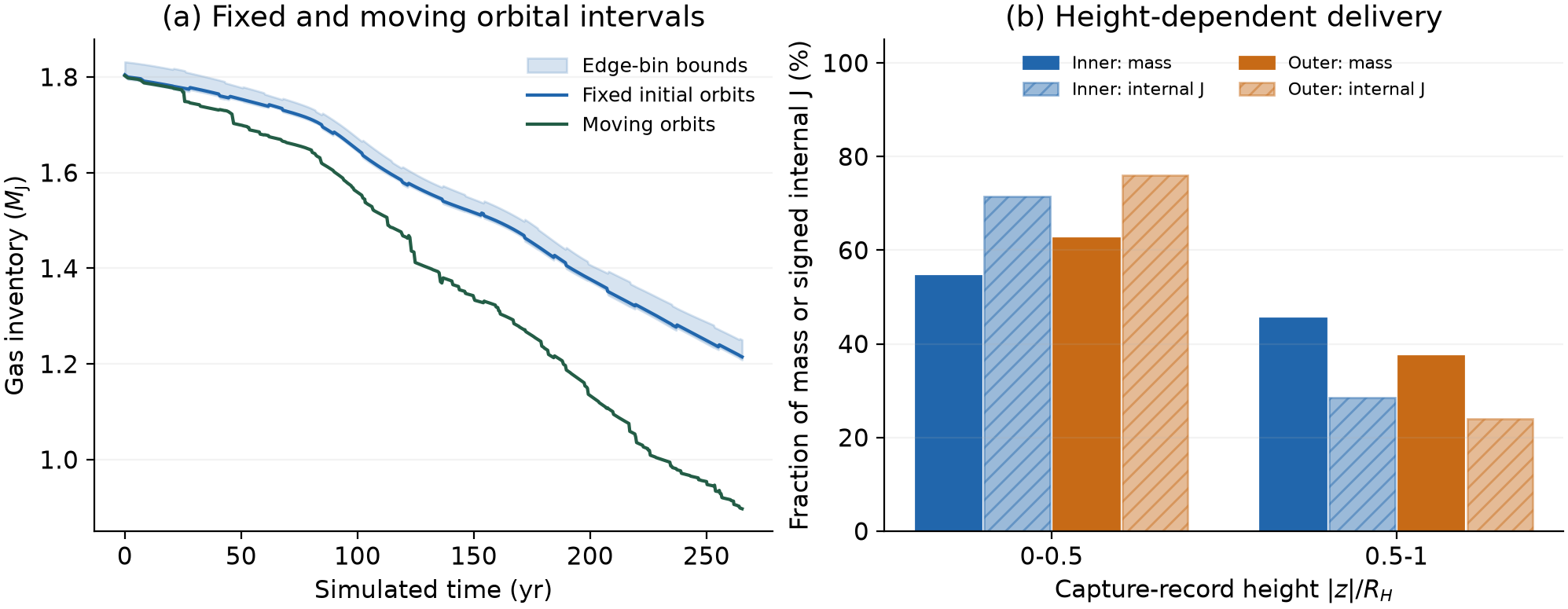}
\caption{Inventory reclassification and height-resolved capture. (a) Gas between fixed post-initialization orbital radii and between the moving orbits. The shaded band brackets the mass in the two unresolved radial edge bins; the fixed-orbit line divides those bins by annular area. (b) Distribution of subsequent captured mass (solid bars) and signed advected internal angular momentum (hatched bars) over capture-record height, each normalized separately for each planet. Height is measured at entry in units of the instantaneous Hill radius. Initial absorption is excluded, azimuth bins are summed, and both disk halves are included once. No capture is recorded in the bins above one Hill radius, as expected for the absorbing Hill-surface geometry. These are entry heights, not launch heights or parcel trajectories.}\label{fig:3}
\end{figure*}

Figure~\ref{fig:3}(a) evaluates both terms from the saved 72-bin radial mass histograms of native exterior gas; gas inside the excluded capture volumes is not reconstructed. Bins lying entirely within the fixed initial interval are summed exactly, and the two intersected edge bins set lower and upper bounds. The plotted estimate divides each edge bin by annular area, assuming uniform surface density within it. These bounds describe unresolved spatial allocation, not statistical uncertainty. At the endpoint, \(1.22\,M_{\rm J}\) remains between the initial orbits (bounds \(1.21\)--\(1.25\,M_{\rm J}\)), compared with \(0.897\,M_{\rm J}\) between the moving orbits. Migration therefore reclassifies about \(0.318\,M_{\rm J}\) (bounds \(0.311\)--\(0.353\,M_{\rm J}\)), and net accumulation outside the initial orbits is about \(0.236\,M_{\rm J}\) (bounds \(0.200\)--\(0.242\,M_{\rm J}\)). Together they make up the \(0.554\,M_{\rm J}\) exterior increase. Net accumulation outside the fixed orbits is thus comparable to the \(0.241\,M_{\rm J}\) captured, rather than 2.3 times larger. Both are net stocks compared with cumulative capture; neither is a gross count of gas passing the planets. The inventory between the initial orbits falls by about 32.6\%, compared with 50.2\% between the moving orbits, so reclassification accounts for about 35.1\% of the moving-orbit decrease.

\textbf{Initialization and growth.} Table~\ref{tab:6} lists the realized initial and final states. The bare seed masses are \(0.262\) and \(0.506\,M_{\rm J}\), with initial \(R_H/H\) of 0.998 and 0.993, close to the nominal design of Table~\ref{tab:2}. After initial capture the planets satisfy \(R_H/H=1\) to within \(3\times10^{-6}\), and both remain close to this regime as they grow, ending at \(R_H/H=1.08\) and 1.14.

\begin{table}[tp]
\centering
\caption{Realized initialization and final saved endpoint. Planetary masses include retained CPD gas. Initial Hill-to-height ratios refer to the post-capture state, not the bare inputs; final quantities use cycle 11,912.}\label{tab:6}
\small\setlength{\tabcolsep}{2pt}
\begin{tabular}{@{}lll@{}}
\toprule
\parbox[t]{0.4800\columnwidth}{\raggedright Realized quantity\strut} & \parbox[t]{0.2400\columnwidth}{\centering Inner\strut} & \parbox[t]{0.2400\columnwidth}{\centering Outer\strut} \\[4pt]
\midrule
\parbox[t]{0.4800\columnwidth}{\raggedright Bare seed mass, \(M_{\rm J}\)\strut} & \parbox[t]{0.2400\columnwidth}{\centering 0.262\strut} & \parbox[t]{0.2400\columnwidth}{\centering 0.506\strut} \\[4pt]
\parbox[t]{0.4800\columnwidth}{\raggedright Post-initial-capture mass, \(M_{\rm J}\)\strut} & \parbox[t]{0.2400\columnwidth}{\centering 0.264\strut} & \parbox[t]{0.2400\columnwidth}{\centering 0.515\strut} \\[4pt]
\parbox[t]{0.4800\columnwidth}{\raggedright Final mass, \(M_{\rm J}\)\strut} & \parbox[t]{0.2400\columnwidth}{\centering 0.355\strut} & \parbox[t]{0.2400\columnwidth}{\centering 0.666\strut} \\[4pt]
\parbox[t]{0.4800\columnwidth}{\raggedright Post-initial-capture orbit, au\strut} & \parbox[t]{0.2400\columnwidth}{\centering 12.6\strut} & \parbox[t]{0.2400\columnwidth}{\centering 19.7\strut} \\[4pt]
\parbox[t]{0.4800\columnwidth}{\raggedright Final orbit, au\strut} & \parbox[t]{0.2400\columnwidth}{\centering 13.2\strut} & \parbox[t]{0.2400\columnwidth}{\centering 18.1\strut} \\[4pt]
\parbox[t]{0.4800\columnwidth}{\raggedright Post-initial-capture \(R_H/H\)\strut} & \parbox[t]{0.2400\columnwidth}{\centering 1.00\strut} & \parbox[t]{0.2400\columnwidth}{\centering 1.00\strut} \\[4pt]
\parbox[t]{0.4800\columnwidth}{\raggedright Final \(R_H/H\)\strut} & \parbox[t]{0.2400\columnwidth}{\centering 1.08\strut} & \parbox[t]{0.2400\columnwidth}{\centering 1.14\strut} \\[4pt]
\bottomrule
\end{tabular}
\end{table}

\begin{figure*}[tp]\centering
\includegraphics[width=0.94\textwidth,height=0.50\textheight,keepaspectratio]{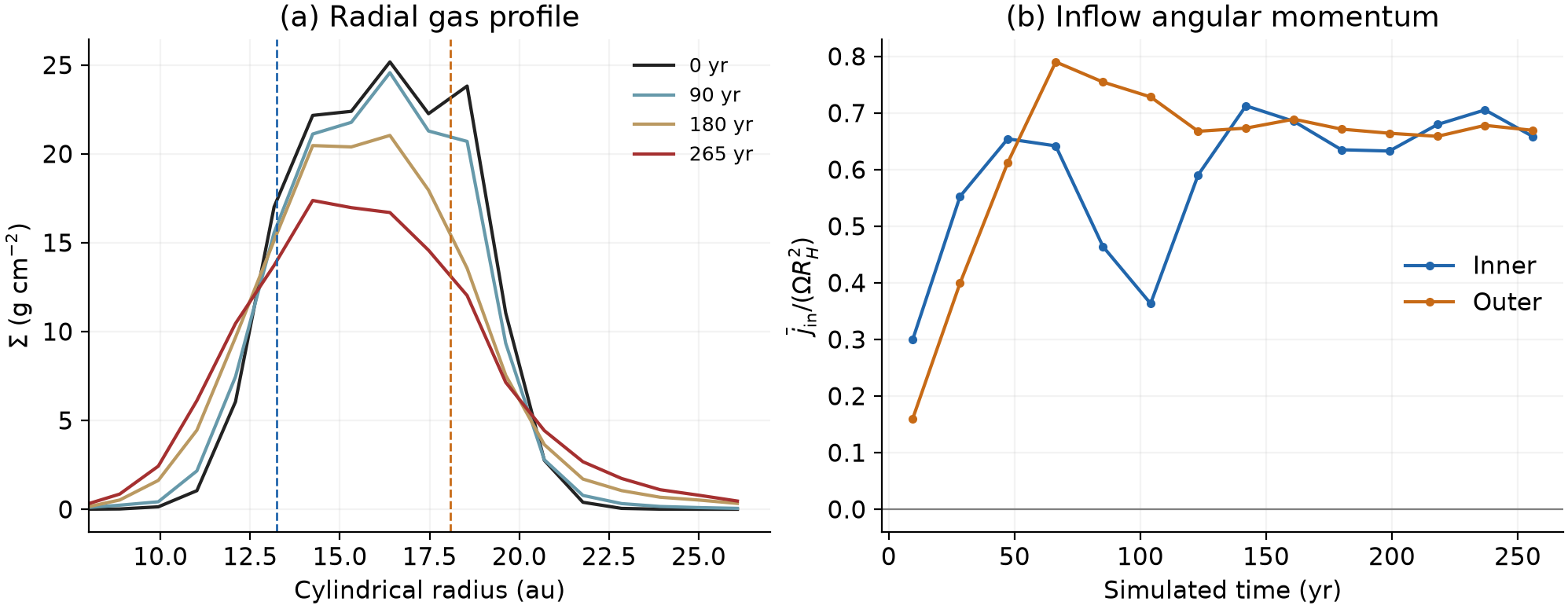}
\caption{Reservoir restructuring and angular-momentum delivery. (a) Azimuthally averaged surface density at four times, integrated over the represented vertical domain and including both disk halves, displayed in 18 radial bins by conservatively combining adjacent saved bins. Excluded capture volumes contain no evolved gas, so the orbital troughs include the absorbing geometry as well as hydrodynamic depletion. Dashed lines mark the final orbital radii. These coarse profiles describe bulk reservoir redistribution, not resolved gap depths. (b) Mass-flux-weighted signed planet-centered angular momentum of newly captured gas in 14 time intervals, normalized by the midpoint orbital frequency and Hill radius. Initial absorption is excluded; pressure impulses are not counted as advected angular momentum.}\label{fig:4}
\end{figure*}

\textbf{Height-dependent delivery.} Figure~\ref{fig:4} shows how the surface-density profile evolves and reports the signed intrinsic angular momentum of captured inflow, \(\widehat j_i=\bar j_{{\rm in},i}/(\Omega_iR_{H,i}^2)\), one of the delivery quantities the companion theory must predict. Each value is accumulated angular momentum divided by accumulated captured mass over an interval, not an unweighted velocity average. No delivery coefficient has been fitted to these measurements.

Figure~\ref{fig:3}(b) resolves capture by entry height. The inner and outer planets receive 45.5\% and 37.4\% of their subsequently captured mass at \(0.5\leq |z|/R_H<1\), and the remainder below \(0.5R_H\). The two height bins are shown without interpolation and describe bulk delivery, not the widths of individual streams. The signed specific internal angular momentum in the upper bin is 0.48 and 0.53 times that in the lower bin, so the upper bin supplies only 28.5\% and 24.0\% of the signed internal capture angular momentum. Height therefore distinguishes angular-momentum delivery as well as mass delivery. The binned mass and internal angular momentum sum to the cumulative capture-ledger changes.

Figures~\ref{fig:2}--\ref{fig:4} and Tables~\ref{tab:5}--\ref{tab:6} use the saved endpoint at cycle 11,912. The global gas-plus-capture-plus-escape mass discrepancy there is below \(10^{-8}\,M_{\rm J}\).

\subsection{A more massive shared reservoir}\label{sec:4.7}

The second simulation increases the initial disk mass from \(2\) to \(6\,M_{\rm J}\) by tripling the density normalization, including the tenuous background of the prescribed profile. The bare seed masses, temperature, \(\Gamma=1.75\) compression, mesh, softening, and boundaries are unchanged. This is a fresh initialization, not a continuation with added gas. Initial capture assigns \(0.0367\,M_{\rm J}\) already enclosed by the Hill spheres to the planets, compared with \(0.0113\,M_{\rm J}\) in the lighter disk. The post-capture inner and outer masses are consequently \(0.268\) and \(0.536\,M_{\rm J}\); their initial \(R_H/H\) values are 1.00 and 1.02. The small initialization-induced orbital changes are retained in the histories rather than reset.

\begin{figure*}[tp]\centering
\includegraphics[width=0.94\textwidth,height=0.50\textheight,keepaspectratio]{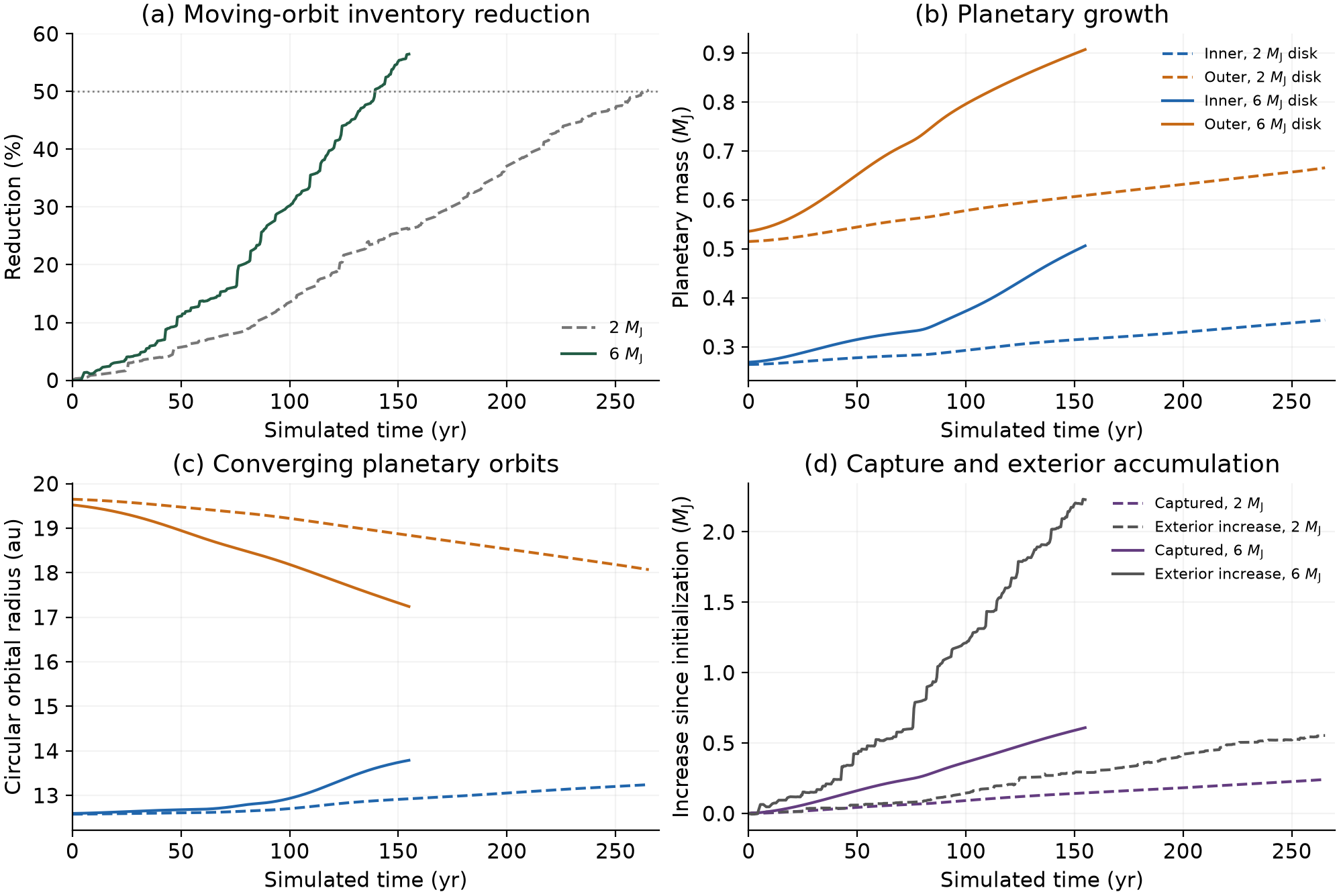}
\caption{Response to a larger shared reservoir. Solid curves show the \(6\,M_{\rm J}\) disk and dashed curves the \(2\,M_{\rm J}\) disk, at the same compression and numerical prescriptions. (a) Interplanetary inventory reduction, relative to each run's post-initial-capture inventory; the dotted line marks 50\%. (b) Inner and outer planetary masses. (c) Their circular orbital radii, using the same planet colors as panel (b). (d) Subsequent capture and the increase of grid gas exterior to the moving planetary orbits. Both exclude their respective initial inventories; the exterior increase includes migration-induced reclassification. Each curve ends at its saved endpoint; the common-age comparison is given in Table~\ref{tab:7}.}\label{fig:5}
\end{figure*}

\textbf{Growth and orbital response.} The \(6\,M_{\rm J}\) run completes in 8.26 hours on eight MPI ranks, advancing to 155 yr. The inner planet grows to \(0.506\,M_{\rm J}\) and the outer to \(0.907\,M_{\rm J}\). After initial capture, their gains are \(0.238\) and \(0.371\,M_{\rm J}\), respectively, so the inner planet receives 39.0\% of subsequent capture. The outer planet remains more massive, but the inner-to-outer planetary mass ratio increases from 0.501 to 0.558. Growth accompanies convergent migration: the inner orbit moves from 12.6 to 13.8 au and the outer from 19.5 to 17.2 au. Their separation decreases from 6.93 to 3.45 au (Figure~\ref{fig:5}).

The physical comparison is clearest at a common age. Table~\ref{tab:7} samples both histories at 150 yr, interpolating between adjacent saved diagnostics. Tripling the disk normalization produces 4.15 times as much subsequent capture by that age, while changing the relative allocation and orbital separation. The response is not simply three times a fixed capture history: the larger gas supply evolves together with the growing, migrating planets. This coupled response is what a delivery model must explain. Because initial capture also changes the planetary masses and orbital radii, this comparison is not a density-power-law measurement at an otherwise identical post-initialization state.

\begin{table}[tp]
\centering
\caption{Reservoir-mass comparison at a common simulated age of 150 yr. Values are linearly interpolated between adjacent saved diagnostics. Subsequent capture excludes gas already enclosed at initialization. Inventory reduction uses each run's own post-initial-capture reference; planetary masses include retained CPD gas.}\label{tab:7}
\small\setlength{\tabcolsep}{2pt}
\begin{tabular}{@{}lll@{}}
\toprule
\parbox[t]{0.4800\columnwidth}{\raggedright Quantity at 150 yr\strut} & \parbox[t]{0.2400\columnwidth}{\centering \(2\,M_{\rm J}\) disk\strut} & \parbox[t]{0.2400\columnwidth}{\centering \(6\,M_{\rm J}\) disk\strut} \\[4pt]
\midrule
\parbox[t]{0.4800\columnwidth}{\raggedright Moving-orbit inventory reduction, \%\strut} & \parbox[t]{0.2400\columnwidth}{\centering 25.8\strut} & \parbox[t]{0.2400\columnwidth}{\centering 55.2\strut} \\[4pt]
\parbox[t]{0.4800\columnwidth}{\raggedright Inner planetary mass, \(M_{\rm J}\)\strut} & \parbox[t]{0.2400\columnwidth}{\centering 0.314\strut} & \parbox[t]{0.2400\columnwidth}{\centering 0.496\strut} \\[4pt]
\parbox[t]{0.4800\columnwidth}{\raggedright Outer planetary mass, \(M_{\rm J}\)\strut} & \parbox[t]{0.2400\columnwidth}{\centering 0.607\strut} & \parbox[t]{0.2400\columnwidth}{\centering 0.899\strut} \\[4pt]
\parbox[t]{0.4800\columnwidth}{\raggedright Subsequent capture, \(M_{\rm J}\)\strut} & \parbox[t]{0.2400\columnwidth}{\centering 0.142\strut} & \parbox[t]{0.2400\columnwidth}{\centering 0.590\strut} \\[4pt]
\parbox[t]{0.4800\columnwidth}{\raggedright Inner share of subsequent capture, \%\strut} & \parbox[t]{0.2400\columnwidth}{\centering 35.5\strut} & \parbox[t]{0.2400\columnwidth}{\centering 38.5\strut} \\[4pt]
\parbox[t]{0.4800\columnwidth}{\raggedright Orbital separation, au\strut} & \parbox[t]{0.2400\columnwidth}{\centering 5.96\strut} & \parbox[t]{0.2400\columnwidth}{\centering 3.58\strut} \\[4pt]
\bottomrule
\end{tabular}
\end{table}

\textbf{Capture, redistribution, and the moving interval.} In the heavier disk, the gas between the planets decreases from \(5.28\) to \(2.30\,M_{\rm J}\), a 56.5\% reduction at the final segment endpoint. The saved diagnostics place the first 50\% crossing near 139 yr, compared with 264 yr in the lighter disk. Because the controller checks the target at segment endpoints, the heavier case continues beyond the first crossing.

Of the original \(6\,M_{\rm J}\), the final grid retains \(2.30\,M_{\rm J}\) between the moving orbits and \(2.91\,M_{\rm J}\) exterior to them. Capture contains \(0.646\,M_{\rm J}\), including initialization, and signed net grid escape is \(0.146\,M_{\rm J}\). As in Section~\ref{sec:4.6}, grid escape is a domain-budget entry. Excluding initialization, the planets capture \(0.609\,M_{\rm J}\), while exterior grid gas increases by \(2.23\,M_{\rm J}\).

As in the lighter disk, part of that exterior increase is reclassification by migration. The fixed-orbit radial-bin estimate leaves \(4.10\,M_{\rm J}\) between the initial orbits, with edge-bin bounds \(4.00\)--\(4.19\,M_{\rm J}\). Approximately \(1.80\,M_{\rm J}\) of the exterior increase is therefore reclassified by migration, and \(0.432\,M_{\rm J}\) is net accumulation exterior to the initial orbits; their bounds are \(1.70\)--\(1.89\) and \(0.342\)--\(0.527\,M_{\rm J}\), respectively. Thus the fixed-region net exterior accumulation is smaller than subsequent capture, despite the much larger exterior increase measured against the moving orbits. The faster reduction of the moving-orbit inventory cannot be identified with faster torque-driven clearing alone. The fixed-initial-orbit inventory decreases by approximately 22.5\%, versus 56.5\% in the moving interval. Migration reclassification supplies about 60.2\% of the moving-interval decrease, compared with 35.1\% in the lighter disk. The two runs therefore illustrate both gas removal and the changing orbital boundaries of the feeding region.

\textbf{Vertically extended delivery persists.} The inner and outer planets receive 42.9\% and 38.0\% of subsequent captured mass at \(0.5\leq |z|/R_H<1\), compared with 45.5\% and 37.4\% in the lighter disk. In the heavier case, the signed specific internal angular momentum in that upper height bin is 0.48 and 0.52 times its value below \(0.5R_H\). Its upper-bin fractions of signed internal capture angular momentum are 26.4\% and 24.2\%, inner then outer. Thus across both PDS disks this upper component delivers 37--46\% of the mass but only 24--29\% of the signed internal angular momentum. These fractions are computed from the cumulative post-initial-capture records, not from rounded displayed ratios. The final \(R_H/H\) values are 1.19 and 1.29. Delivery therefore remains vertically extended during substantial growth, and its angular momentum depends on entry height. These measurements provide a direct target for the meridional delivery theory, which is not fitted here.

\textbf{Physical diagnostic and geometric evolution window.} Integrating the prescribed initial vertical density over the represented disk thickness gives a minimum stellar-Keplerian estimate \(Q_{\rm K}=c_s\Omega_K/(\pi G\Sigma)\) of approximately 5.47 in the lighter disk and 1.82 in the heavier disk, near 18.1 au. Here \(\Omega_K=(GM_\star/R^3)^{1/2}\), and \(\Sigma\) includes both hemispheres before initial capture. These are initializer estimates of the physical diagnostic discussed in Section~\ref{sec:2.7}, not measurements of the evolved disk's gravitational stability.

At the heavier run's endpoint the physical Hill radii sum to 3.62 au, whereas the radial orbital separation is 3.45 au. Their azimuthal separation is then 1.69 radians, so the capture spheres do not overlap at that endpoint. At conjunction with these same masses and orbital radii, however, their surfaces would overlap by approximately 0.172 au, even before numerical buffers. Growth and convergent migration have thus consumed the initial conjunction clearance. This limits continuation with separate capture regions; it is not evidence of an overlap in the reported endpoint.

\subsection{Scaled Solar-System result}\label{sec:4.8}

The \(\Gamma=3\), \(2\,M_{\rm J}\) Solar-System calculation described in Section~\ref{sec:3.2} follows 30.2 yr of coupled growth and migration in 13.6 hours of production execution on eight MPI ranks (Figure~\ref{fig:6}; run record in Appendix~\ref{sec:A.5}). The root mesh is \(36\times6\times112\), with one additional refinement level; the finest equivalent mesh is \(72\times12\times224\). The initial inner and outer planets have approximately 2.51 and 6.27 radial cells per \(H=R_H\), respectively. This is a deliberately coarse simulation using the same perfect-capture treatment as the PDS calculations. Appendix~\ref{sec:A.4} gives its own dimensional and code-unit numerical parameters. At the last diagnostic the inner and outer \(R_H/H\) values are 1.28 and 1.50, respectively, corresponding to thermal-mass parameters \(q_{\rm th}=3(R_H/H)^3\) of 6.26 and 10.1. The outer planet thus reaches the vicinity of the strongly superthermal transition at \(q_{\rm th}\sim10\) identified by \citet{ref2}. The growing Hill sphere samples an increasing fraction of the disk's vertical thickness. This extends the delivery-model test from the initial \(R_H=H\) configuration toward a regime with a different dependence of gas supply on mass and scale height. The thermal-mass parameter is not held fixed during growth.

After initial capture, the inner and outer planetary masses are \(0.0264\) and \(0.0646\,M_{\rm J}\). They reach \(0.0617\) and \(0.171\,M_{\rm J}\), with subsequent gains of \(0.0352\) and \(0.106\,M_{\rm J}\). The outer planet receives 75.0\% of this subsequent capture. Their orbital radii change from 1.74 and 3.16 au to 1.87 and 2.67 au: the separation decreases by 43.6\%, from 1.43 to 0.805 au. The starting radii here follow initial capture. Table~\ref{tab:3} instead lists the nominal input radii, 1.73 and 3.19 au: assigning the initially enclosed gas and its angular momentum to each planet shifts its circular orbit before the subsequent evolution begins. These shifts are retained, not reset.

\textbf{Reservoir evolution.} Gas between the moving orbits decreases from \(1.82\) to \(0.916\,M_{\rm J}\), approximately half its post-initial-capture inventory. The binned profile gives a much smaller reduction, approximately 14\%, within the fixed initial orbital interval. The precise fixed-region value is limited by the two radial edge bins, as shown in Figure~\ref{fig:6}. Convergent migration therefore contributes strongly to the reduction of the interplanetary inventory. Grid gas exterior to the moving planetary orbits increases from \(0.174\) to \(0.888\,M_{\rm J}\). As in the PDS runs, this change combines redistribution relative to the original orbits with reclassification as those orbits move. Of the original \(2\,M_{\rm J}\) disk, \(1.80\,M_{\rm J}\) remains on the grid, \(0.147\,M_{\rm J}\) has been captured, and \(0.0488\,M_{\rm J}\) has left through the computational boundaries. The captured total includes \(0.00570\,M_{\rm J}\) initially inside the Hill spheres.

\textbf{Endpoint and geometric limit.} The last common saved diagnostic is cycle 12,160, at a 49.7\% inventory reduction. The run subsequently stopped at cycle 12,172 when a coordinate cell intersected both buffered managed sink patches. This guard concerns the numerical patches, not a demonstrated overlap of the physical Hill spheres. We report this endpoint without changing the capture-patch geometry. The separate-patch geometry ultimately limits the available evolution window as capture and migration bring the planets together.

\begin{figure*}[tp]\centering
\includegraphics[width=0.94\textwidth,height=0.50\textheight,keepaspectratio]{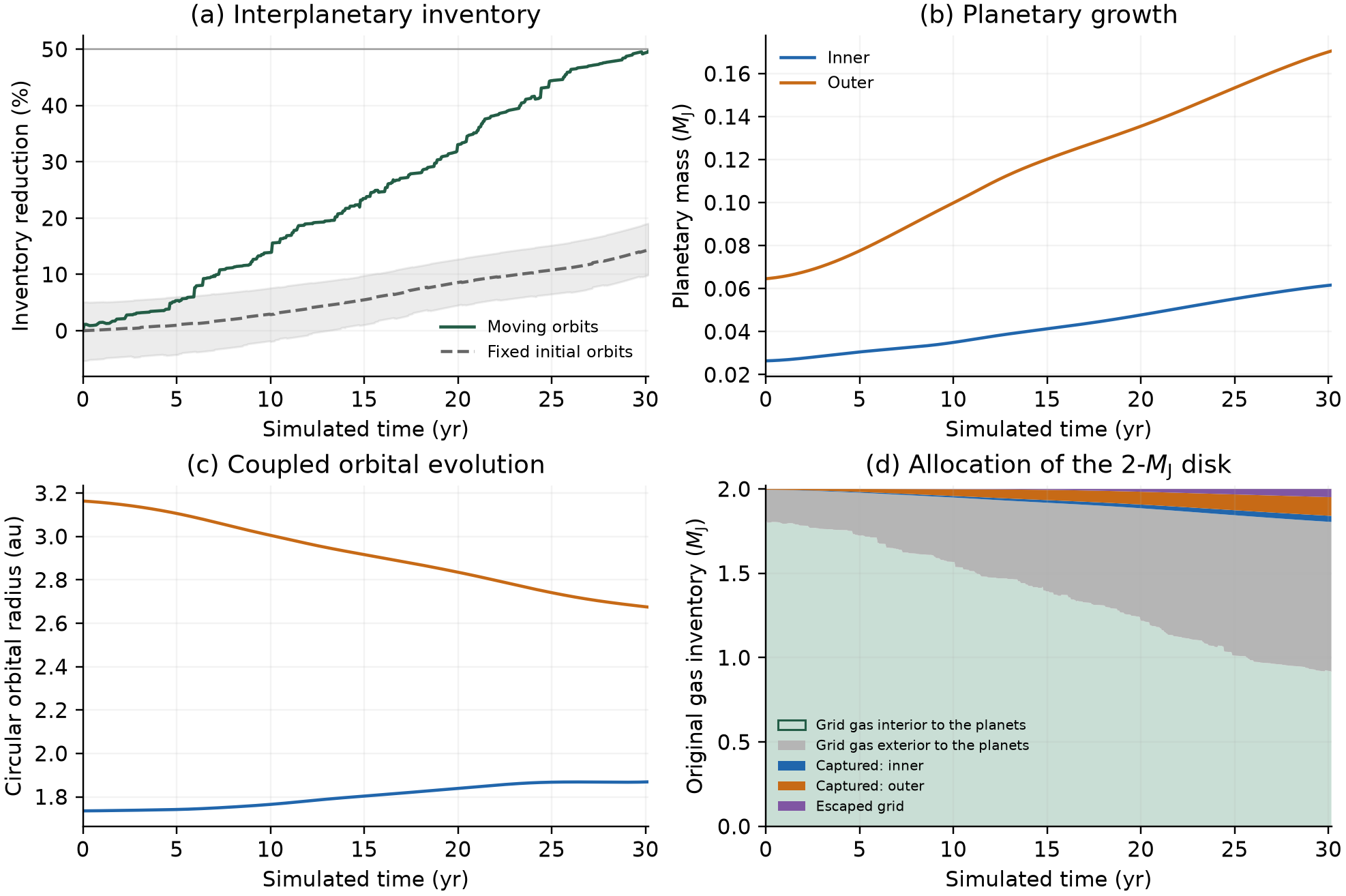}
\caption{Scaled Solar-System evolution at \(\Gamma=3\). (a) Reduction of gas between the moving orbits and within the fixed initial orbital interval, each relative to its own post-initial-capture inventory. The fixed-region estimate splits radial edge bins by annular area; shading brackets the unresolved edge-bin contributions at the initial and current times. The horizontal line marks 50\%. (b) Inner and outer planetary masses. (c) Their circular orbital radii. (d) Partition of the original disk, including initial capture. Green denotes gas between the moving orbits; gray includes gas inside the inner orbit and beyond the outer orbit. Gray is not boundary escape. All panels end at the last common saved diagnostic before the managed-patch guard stopped the run.}\label{fig:6}
\end{figure*}

\section{Angular-momentum budgets and convergent migration}\label{sec:5}

\subsection{Total angular-momentum conservation}\label{sec:5.1}

Table~\ref{tab:8} accounts for angular momentum retained in the gas, transferred to the two planets, and exported through the computational boundaries. It compares the two PDS reservoirs of Sections~\ref{sec:4.6}--\ref{sec:4.7} and the scaled Solar-System calculation of Section~\ref{sec:4.8}. The columns describe each run's retained endpoint, not matched physical times or identical reduction fractions.

Let \(L_0\) be each system's initial star-centered angular momentum. The reported residual is \begin{equation}\label{eq:32}
\fitmath{\delta_L=\frac{L_{g,z}+\sum_i(L_{o,i}+J_i)
                    +L_{\rm boundary,out}-L_0}{L_0}.}
\end{equation} The gas inventory includes material on both sides of the planetary orbits, whether or not it remains available for capture. Boundary export is signed and counted once for the shared gas. Capture and gravitational or pressure torques redistribute angular momentum internally; they are not additional terms to add to this total.

\begin{table*}[tp]
\centering
\caption{Star-centered angular-momentum changes from each post-initial-capture state, normalized by that system's initial total angular momentum. The two PDS disks use saved cycles 11,912 (265 yr) and 6,352 (155 yr); the Solar-System column uses the cycle-12,160 diagnostic (30.2 yr). These are separate endpoints, not a matched-time comparison. Native checkpoint inventories are available at the two PDS endpoints, not at the later Solar-System diagnostic. Residuals use unrounded records; no clipping contribution is subtracted.}\label{tab:8}
\small\setlength{\tabcolsep}{2pt}
\begin{tabular}{@{}llll@{}}
\toprule
\parbox[t]{0.2600\textwidth}{\raggedright Contribution\strut} & \parbox[t]{0.2300\textwidth}{\centering PDS: \(2\,M_{\rm J}\) disk\strut} & \parbox[t]{0.2300\textwidth}{\centering PDS: \(6\,M_{\rm J}\) disk\strut} & \parbox[t]{0.2300\textwidth}{\centering Solar: \(2\,M_{\rm J}\) disk\strut} \\[4pt]
\midrule
\parbox[t]{0.2600\textwidth}{\raggedright Gas remaining on grid\strut} & \parbox[t]{0.2300\textwidth}{\centering \(-1.28\times10^{-1}\)\strut} & \parbox[t]{0.2300\textwidth}{\centering \(-1.12\times10^{-1}\)\strut} & \parbox[t]{0.2300\textwidth}{\centering \(-9.27\times10^{-2}\)\strut} \\[4pt]
\parbox[t]{0.2600\textwidth}{\raggedright Inner orbital ledger\strut} & \parbox[t]{0.2300\textwidth}{\centering \(+3.21\times10^{-2}\)\strut} & \parbox[t]{0.2300\textwidth}{\centering \(+3.45\times10^{-2}\)\strut} & \parbox[t]{0.2300\textwidth}{\centering \(+1.51\times10^{-2}\)\strut} \\[4pt]
\parbox[t]{0.2600\textwidth}{\raggedright Outer orbital ledger\strut} & \parbox[t]{0.2300\textwidth}{\centering \(+4.94\times10^{-2}\)\strut} & \parbox[t]{0.2300\textwidth}{\centering \(+5.21\times10^{-2}\)\strut} & \parbox[t]{0.2300\textwidth}{\centering \(+5.01\times10^{-2}\)\strut} \\[4pt]
\parbox[t]{0.2600\textwidth}{\raggedright Inner internal ledger\strut} & \parbox[t]{0.2300\textwidth}{\centering \(+1.37\times10^{-4}\)\strut} & \parbox[t]{0.2300\textwidth}{\centering \(+1.49\times10^{-4}\)\strut} & \parbox[t]{0.2300\textwidth}{\centering \(+4.36\times10^{-5}\)\strut} \\[4pt]
\parbox[t]{0.2600\textwidth}{\raggedright Outer internal ledger\strut} & \parbox[t]{0.2300\textwidth}{\centering \(+4.58\times10^{-4}\)\strut} & \parbox[t]{0.2300\textwidth}{\centering \(+5.18\times10^{-4}\)\strut} & \parbox[t]{0.2300\textwidth}{\centering \(+3.59\times10^{-4}\)\strut} \\[4pt]
\parbox[t]{0.2600\textwidth}{\raggedright Boundary export\strut} & \parbox[t]{0.2300\textwidth}{\centering \(+4.58\times10^{-2}\)\strut} & \parbox[t]{0.2300\textwidth}{\centering \(+2.50\times10^{-2}\)\strut} & \parbox[t]{0.2300\textwidth}{\centering \(+2.71\times10^{-2}\)\strut} \\[4pt]
\parbox[t]{0.2600\textwidth}{\raggedright Raw residual \(\delta_L\)\strut} & \parbox[t]{0.2300\textwidth}{\centering \(+1.57\times10^{-9}\)\strut} & \parbox[t]{0.2300\textwidth}{\centering \(+1.90\times10^{-9}\)\strut} & \parbox[t]{0.2300\textwidth}{\centering \(+3.65\times10^{-8}\)\strut} \\[4pt]
\bottomrule
\end{tabular}
\end{table*}

For the two PDS endpoints, independent integration of the native checkpoint cells reproduces the gas inventory. The individual planetary orbital and internal exchanges close to roundoff, with absolute residuals below \(2.4\times10^{-17}\) in code angular-momentum units. The complete budgets close to approximately two parts per billion. This net global residual, reported without any numerical-source subtraction, is neither a bound on accumulated absolute clipping nor proof of capture-rate convergence.

For the Solar-System diagnostic endpoint in Section~\ref{sec:4.8}, the same gas-plus-planets-plus-boundary sum has a fractional raw angular-momentum residual of \(+3.65\times10^{-8}\). The corresponding mass discrepancy is \(9.21\times10^{-8}\,M_{\rm J}\). These small global discrepancies are reported without correction. Unlike the two PDS endpoints in Table~\ref{tab:8}, this later Solar-System diagnostic does not have a matching full checkpoint for independent native-cell integration; the last restart checkpoint is earlier. The possible ceiling contribution discussed below also applies to this run, without establishing a complete attribution.

In the Solar-System run the gas loses 9.27\% of \(L_0\), while the inner and outer orbital ledgers gain 1.51\% and 5.01\%, respectively. The retained internal CPD ledgers gain a further 0.0402\% in total, and signed boundary export carries 2.71\%. These are transfers relative to the post-initial-capture state, not final inventory fractions. Angular momentum retained by gas outside the planetary orbits remains in the gas term; it is not counted as boundary loss. The two local orbital and internal exchange identities in Section~\ref{sec:5.2} close with absolute residuals below \(8\times10^{-18}\) in code angular-momentum units. This local closure and the small global residual are distinct measurements.

\textbf{Possible contribution from the velocity ceiling.} The component-wise ceiling described in Section~\ref{sec:2.5} modifies stored angular momentum when it clips the azimuthal velocity. For an evolved cell of mass \(m_c\) and discrete cylindrical lever arm \(\mathcal R_c\), let \(v_{\phi,c}\) and \(v_{\phi,c}^{\rm clip}\) denote the azimuthal velocities before and after clipping. The direct change is \begin{equation}\label{eq:33}
\fitmath{\Delta L_{z,c}^{\rm clip}
   =m_c\mathcal R_c\bigl(v_{\phi,c}^{\rm clip}-v_{\phi,c}\bigr).}
\end{equation} Radial or polar clipping alone does not directly change \(L_z\). In low-density atmospheric gas, the small cell mass can make this angular-momentum change tiny even for a substantial velocity adjustment. There is no compensating planetary transfer or boundary-export entry for this operation, so it is a plausible contributor to the residual. Its sign depends on the clipped azimuthal velocity; clipping need not produce a net loss of signed angular momentum. The cumulative contribution to accepted updates was not separately recorded, so the measured residual is not attributed to clipping alone.

\subsection{Partition between the retained CPD and the circular orbit}\label{sec:5.2}

Captured gas does not disappear from the gravitating system. Its mass is added to the retained CPD ledger, so the mass used for the planetary potential and the circular orbit is \(M_i=M_{{\rm seed},i}+M_{{\rm CPD},i}\). The seed mass itself is fixed: no transport from the retained CPD onto a planetary surface is prescribed in these simulations. Nor is the interior of the capture sphere evolved hydrodynamically.

These CPD assumptions constitute a pragmatic implementation approach, not a developed physical model of a circumplanetary disk. They retain the captured mass and angular momentum and couple them to the planet's gravity and orbital evolution, without prescribing the disk's internal structure, thermal evolution, or angular-momentum transport. The purpose is to follow delivery and its orbital consequences; transporting the retained gas onto the planetary surface requires further modeling.

For each captured contribution of mass \(\delta m\), the code retains the star-centered angular momentum \(\delta L_\star\) removed by the conservative gas update. It separately measures the signed angular momentum about the moving planet, \begin{equation}\label{eq:34}
\fitmath{\delta J=\delta m\,[(\mathbf r-\mathbf r_i)
                    \times(\mathbf v-\mathbf v_i)]_z.}
\end{equation} This contribution enters the internal CPD ledger; the remainder \(\delta L_\star-\delta J\) enters the orbital ledger. Thus the CPD carries both its internal angular momentum and its share of the planet's orbital motion: assigning all captured star-centered angular momentum to the orbit would count the internal contribution incorrectly. The circularization radius is a delivery diagnostic, not a second sink or a resolved redistribution of gas within the excluded volume.

Let \(C_{L,i}\) and \(C_{J,i}\) be the accumulated captured star-centered and internal angular momenta after initialization. Let \(G_i\) be the cumulative gravitational angular impulse received by the planet: the sum of the single-update transfers \(-I_{g,i}\) defined in Section~\ref{sec:4.2}. Likewise, \(P_{L,i}\) and \(P_{J,i}\) are the cumulative star-centered and internal pressure impulses, obtained by summing \(I_{P,L,i}\) and \(I_{P,J,i}\) over the same accepted updates. Their signs are positive when they add angular momentum to that ledger. Over the saved evolution, \begin{equation}\label{eq:35}
\fitmath{\begin{aligned}
 \Delta J_i &= C_{J,i}+P_{J,i},\\
 \Delta L_{o,i} &= C_{L,i}-C_{J,i}+G_i+P_{L,i}-P_{J,i}.
 \end{aligned}}
\end{equation} Gravity is paired with the equal-and-opposite torque on the exterior gas. Pressure traction at the absorbing interface is likewise paired with the gas update. Its small internal contribution from the faceted interface is retained in \(J_i\), not discarded or treated as physical CPD transport. Adding the two equations recovers the full star-centered transfer without double counting. After each accepted exchange the code recomputes \(a_i=(L_{o,i}/M_i)^2/(GM_\star)\), retaining a circular orbit.

The gravitational torque on each planet can in turn be separated by the gas's position relative to that planet's own orbit. With \(R\) the star-centered cylindrical radius and \(g_{\phi,i}\) the azimuthal acceleration of gas due to planet \(i\), define \begin{equation}\label{eq:36}
\fitmath{\begin{aligned}
 \tau_{g,i}^{<}&=-\int_{R<a_i}\rho R g_{\phi,i}\,dV,\\
 \tau_{g,i}^{>}&=-\int_{R\geq a_i}\rho R g_{\phi,i}\,dV,\\
 G_i&=G_i^{<}+G_i^{>},\\
 G_i^{\lessgtr}&=\int_{0^+}^{t}\tau_{g,i}^{\lessgtr}\,dt'.
 \end{aligned}}
\end{equation} The integrals exclude both capture volumes. The starward (\(R<a_i\)) and outward (\(R\geq a_i\)) contributions are defined relative to the instantaneous orbit of the planet exerting the force, not to the pair-wide inventory labels. Gas between the planets lies outward of the inner orbit and starward of the outer orbit. The attribution does not depend on which planet eventually captures that gas. No sign is imposed on either one-sided contribution. The saved cumulative ledgers retain \(G_i\), not \(G_i^{<}\) and \(G_i^{>}\) separately, so Table~\ref{tab:9} reports their measured sum. An instantaneous split reconstructed from a checkpoint would not recover the missing time-integrated one-sided budgets.

\subsection{PDS endpoint angular-momentum excess budgets}\label{sec:5.3}

Both planets gain total orbital angular momentum as they gain mass, but that fact alone does not determine the direction of migration. The relevant quantity is orbital angular momentum per unit planetary mass. For an exact endpoint budget, define \(\ell_{i,0}=L_{o,i}(0^+)/M_i(0^+)\), fixed at the post-initial-capture state. With \(\ell_i=L_{o,i}/M_i\), define \begin{equation}\label{eq:37}
\fitmath{\begin{aligned}
 \mathcal E_i &\equiv C_{L,i}-C_{J,i}-\ell_{i,0}\Delta M_i,\\
 \mathcal Q_i &\equiv M_i(t)[\ell_i(t)-\ell_{i,0}]\\
     &=\mathcal E_i+G_i+P_{L,i}-P_{J,i}.
 \end{aligned}}
\end{equation} The term \(\ell_{i,0}\Delta M_i\) is the angular momentum needed to add the captured mass at unchanged initial orbital specific angular momentum; it is a reference subtraction, not a physical loss. \(\mathcal E_i\) measures the captured excess or deficit relative to that value. Positive \(\mathcal Q_i\) means outward migration and negative \(\mathcal Q_i\) inward migration. This decomposition uses cumulative exchanges and the measured mass gain, without freezing the orbit during the simulation or introducing an independent migration coefficient.

\begin{table*}[tp]
\centering
\caption{Capture and torque contributions to each planet's orbital budget, from initialization to the same endpoints as Table~\ref{tab:8}. All entries are in $10^{-3}L_0$, with each disk's own initial system angular momentum $L_0$. Capture excludes initial absorption. The captured orbital term plus gravitational and orbital-pressure impulses gives $\Delta L_o$. Subtracting the mass reference gives $\mathcal Q=\mathcal E+G+P_L-P_J$. The internal capture entry is already included in the captured star-centered entry and must not be added to it. The gravitational row is the measured sum of interior and exterior contributions, not separate one-sided measurements. Rounding can affect displayed sums.}\label{tab:9}
\small\setlength{\tabcolsep}{2pt}
\begin{tabular}{@{}lllll@{}}
\toprule
\parbox[t]{0.2600\textwidth}{\raggedright Contribution, in \(10^{-3}L_0\)\strut} & \parbox[t]{0.1725\textwidth}{\centering \(2\,M_{\rm J}\): inner\strut} & \parbox[t]{0.1725\textwidth}{\centering \(2\,M_{\rm J}\): outer\strut} & \parbox[t]{0.1725\textwidth}{\centering \(6\,M_{\rm J}\): inner\strut} & \parbox[t]{0.1725\textwidth}{\centering \(6\,M_{\rm J}\): outer\strut} \\[4pt]
\midrule
\parbox[t]{0.2600\textwidth}{\raggedright Captured star-centered, \(C_L\)\strut} & \parbox[t]{0.1725\textwidth}{\centering \(+30.9\)\strut} & \parbox[t]{0.1725\textwidth}{\centering \(+54.1\)\strut} & \parbox[t]{0.1725\textwidth}{\centering \(+33.8\)\strut} & \parbox[t]{0.1725\textwidth}{\centering \(+55.3\)\strut} \\[4pt]
\parbox[t]{0.2600\textwidth}{\raggedright Retained internal capture, \(C_J\)\strut} & \parbox[t]{0.1725\textwidth}{\centering \(+0.137\)\strut} & \parbox[t]{0.1725\textwidth}{\centering \(+0.458\)\strut} & \parbox[t]{0.1725\textwidth}{\centering \(+0.149\)\strut} & \parbox[t]{0.1725\textwidth}{\centering \(+0.518\)\strut} \\[4pt]
\parbox[t]{0.2600\textwidth}{\raggedright Capture assigned to orbit, \(C_L-C_J\)\strut} & \parbox[t]{0.1725\textwidth}{\centering \(+30.8\)\strut} & \parbox[t]{0.1725\textwidth}{\centering \(+53.7\)\strut} & \parbox[t]{0.1725\textwidth}{\centering \(+33.6\)\strut} & \parbox[t]{0.1725\textwidth}{\centering \(+54.8\)\strut} \\[4pt]
\parbox[t]{0.2600\textwidth}{\raggedright Gravitational impulse, \(G^{<}+G^{>}\)\strut} & \parbox[t]{0.1725\textwidth}{\centering \(+1.34\)\strut} & \parbox[t]{0.1725\textwidth}{\centering \(-4.52\)\strut} & \parbox[t]{0.1725\textwidth}{\centering \(+0.732\)\strut} & \parbox[t]{0.1725\textwidth}{\centering \(-2.77\)\strut} \\[4pt]
\parbox[t]{0.2600\textwidth}{\raggedright Orbital pressure impulse, \(P_L-P_J\)\strut} & \parbox[t]{0.1725\textwidth}{\centering \(-0.0324\)\strut} & \parbox[t]{0.1725\textwidth}{\centering \(+0.231\)\strut} & \parbox[t]{0.1725\textwidth}{\centering \(+0.163\)\strut} & \parbox[t]{0.1725\textwidth}{\centering \(+0.117\)\strut} \\[4pt]
\parbox[t]{0.2600\textwidth}{\raggedright Total orbital change, \(\Delta L_o\)\strut} & \parbox[t]{0.1725\textwidth}{\centering \(+32.1\)\strut} & \parbox[t]{0.1725\textwidth}{\centering \(+49.4\)\strut} & \parbox[t]{0.1725\textwidth}{\centering \(+34.5\)\strut} & \parbox[t]{0.1725\textwidth}{\centering \(+52.1\)\strut} \\[4pt]
\parbox[t]{0.2600\textwidth}{\raggedright Mass reference, \(\ell_0\Delta M\)\strut} & \parbox[t]{0.1725\textwidth}{\centering \(+29.1\)\strut} & \parbox[t]{0.1725\textwidth}{\centering \(+60.3\)\strut} & \parbox[t]{0.1725\textwidth}{\centering \(+31.4\)\strut} & \parbox[t]{0.1725\textwidth}{\centering \(+61.1\)\strut} \\[4pt]
\parbox[t]{0.2600\textwidth}{\raggedright Accretion excess or deficit, \(\mathcal E\)\strut} & \parbox[t]{0.1725\textwidth}{\centering \(+1.65\)\strut} & \parbox[t]{0.1725\textwidth}{\centering \(-6.66\)\strut} & \parbox[t]{0.1725\textwidth}{\centering \(+2.20\)\strut} & \parbox[t]{0.1725\textwidth}{\centering \(-6.33\)\strut} \\[4pt]
\parbox[t]{0.2600\textwidth}{\raggedright Net migration budget, \(\mathcal Q\)\strut} & \parbox[t]{0.1725\textwidth}{\centering \(+2.96\)\strut} & \parbox[t]{0.1725\textwidth}{\centering \(-11.0\)\strut} & \parbox[t]{0.1725\textwidth}{\centering \(+3.10\)\strut} & \parbox[t]{0.1725\textwidth}{\centering \(-8.99\)\strut} \\[4pt]
\bottomrule
\end{tabular}
\end{table*}

Table~\ref{tab:9} shows that accretion supplies the larger term in this endpoint decomposition of convergence in both simulations. The inner planet captures an excess of orbital angular momentum per unit added mass and moves outward; the outer planet captures a deficit and moves inward. This is the expected sense for feeding from the shared reservoir between the planets. The net gravitational-plus-pressure impulses reinforce that convergence, but are smaller than the accretion contribution. In the \(6\,M_{\rm J}\) disk, \(|\mathcal E_i|/|G_i+P_{L,i}-P_{J,i}|\) is 2.46 for the inner planet and 2.38 for the outer planet. Accretion accounts for 71\% and 70\% of their respective \(\mathcal Q_i\); the corresponding fractions in the \(2\,M_{\rm J}\) disk are 56\% and 61\%. These are shares of the endpoint excess \(\mathcal Q_i\), not an attribution of the accumulated change in orbital specific angular momentum to each process. In particular, \(\mathcal E_i\) includes mass accreted at an orbital specific angular momentum already changed by earlier torques. Nor do these shares describe a counterfactual evolution with the torques removed. The internal pressure contributions are less than \(10^{-7}L_0\) in all four ledgers and are included throughout.

The inward motion of the outer planet despite its positive \(\Delta L_o\) is particularly instructive. Its orbital angular momentum grows, but more slowly than in proportion to its mass, which is what maintaining the initial orbit would require. Preserving the CPD--orbital split and the captured specific angular momentum is therefore essential to recovering the convergent evolution. Gravitational torques alone do not explain it.

\subsection{Solar-System accretion and migration budget}\label{sec:5.4}

Table~\ref{tab:10} applies the same decomposition to the Solar-System endpoint, excluding gas absorbed at initialization. Both planets gain orbital angular momentum, but their specific angular momenta change in opposite directions. The inner planet's captured orbital angular momentum exceeds the mass reference \(\ell_{i,0}\Delta M_i\), while the outer planet's falls short of it. Thus capture contributes outward motion of the inner planet and inward motion of the outer, narrowing the shared feeding region.

\begin{table}[tp]
\centering
\caption{Solar-System capture and torque contributions through cycle 12,160 (30.2 yr), in $10^{-3}L_0$. Definitions are identical to Table~\ref{tab:9}, using this system's own initial angular momentum. Capture excludes initialization; internal capture is part of the star-centered entry, not an additional transfer. The mass reference distinguishes orbital-angular-momentum growth from migration. All differences and endpoint shares are calculated before rounding; the displayed capture terms can coincide despite their small internal difference.}\label{tab:10}
\small\setlength{\tabcolsep}{2pt}
\begin{tabular}{@{}lll@{}}
\toprule
\parbox[t]{0.4800\columnwidth}{\raggedright Contribution, in \(10^{-3}L_0\)\strut} & \parbox[t]{0.2400\columnwidth}{\centering Solar: inner\strut} & \parbox[t]{0.2400\columnwidth}{\centering Solar: outer\strut} \\[4pt]
\midrule
\parbox[t]{0.4800\columnwidth}{\raggedright Captured star-centered, \(C_L\)\strut} & \parbox[t]{0.2400\columnwidth}{\centering \(+15.0\)\strut} & \parbox[t]{0.2400\columnwidth}{\centering \(+52.0\)\strut} \\[4pt]
\parbox[t]{0.4800\columnwidth}{\raggedright Retained internal capture, \(C_J\)\strut} & \parbox[t]{0.2400\columnwidth}{\centering \(+0.0436\)\strut} & \parbox[t]{0.2400\columnwidth}{\centering \(+0.359\)\strut} \\[4pt]
\parbox[t]{0.4800\columnwidth}{\raggedright Capture assigned to orbit, \(C_L-C_J\)\strut} & \parbox[t]{0.2400\columnwidth}{\centering \(+15.0\)\strut} & \parbox[t]{0.2400\columnwidth}{\centering \(+51.7\)\strut} \\[4pt]
\parbox[t]{0.4800\columnwidth}{\raggedright Gravitational impulse, \(G^{<}+G^{>}\)\strut} & \parbox[t]{0.2400\columnwidth}{\centering \(+0.184\)\strut} & \parbox[t]{0.2400\columnwidth}{\centering \(-1.72\)\strut} \\[4pt]
\parbox[t]{0.4800\columnwidth}{\raggedright Orbital pressure impulse, \(P_L-P_J\)\strut} & \parbox[t]{0.2400\columnwidth}{\centering \(-0.0183\)\strut} & \parbox[t]{0.2400\columnwidth}{\centering \(+0.150\)\strut} \\[4pt]
\parbox[t]{0.4800\columnwidth}{\raggedright Total orbital change, \(\Delta L_o\)\strut} & \parbox[t]{0.2400\columnwidth}{\centering \(+15.1\)\strut} & \parbox[t]{0.2400\columnwidth}{\centering \(+50.1\)\strut} \\[4pt]
\parbox[t]{0.4800\columnwidth}{\raggedright Mass reference, \(\ell_0\Delta M\)\strut} & \parbox[t]{0.2400\columnwidth}{\centering \(+14.2\)\strut} & \parbox[t]{0.2400\columnwidth}{\centering \(+57.6\)\strut} \\[4pt]
\parbox[t]{0.4800\columnwidth}{\raggedright Accretion excess or deficit, \(\mathcal E\)\strut} & \parbox[t]{0.2400\columnwidth}{\centering \(+0.774\)\strut} & \parbox[t]{0.2400\columnwidth}{\centering \(-5.90\)\strut} \\[4pt]
\parbox[t]{0.4800\columnwidth}{\raggedright Net migration budget, \(\mathcal Q\)\strut} & \parbox[t]{0.2400\columnwidth}{\centering \(+0.939\)\strut} & \parbox[t]{0.2400\columnwidth}{\centering \(-7.46\)\strut} \\[4pt]
\bottomrule
\end{tabular}
\end{table}

Accretion supplies 82.4\% of the inner planet's positive \(\mathcal Q_i\) and 79.0\% of the outer planet's negative \(\mathcal Q_i\). Gravity reinforces each change, contributing 19.5\% and 23.0\%, while pressure opposes it by 1.95\% and 2.01\%, respectively. Equivalently, the accretion term is 4.68 and 3.76 times the net gravitational-plus-pressure term in magnitude. These signed endpoint shares use the same fixed initial specific-angular-momentum reference as Table~\ref{tab:9}; they are not percentages of radial displacement. The full measured budget recovers radius ratios of 1.08 and 0.845, consistent with the orbital histories in Figure~\ref{fig:6}.

The internal CPD capture contributions remain in \(C_J\) and are subtracted before assigning capture to the orbit. Internal pressure changes of \(-9.97\times10^{-9}L_0\) and \(+4.99\times10^{-8}L_0\), inner then outer, are also retained in the internal ledgers. They are included in the orbital pressure term with the opposite sign. Thus neither the small intrinsic CPD component nor the much larger orbital component is lost or counted twice. These endpoint transfers establish the orbital budget. The distinct step-weighted attribution below determines how capture and torques accumulated the change in orbital specific angular momentum.

\subsection{Step-weighted attribution of orbital evolution}\label{sec:5.5}

The fixed initial reference in \(\mathcal E_i\) is useful for endpoint accounting but does not by itself isolate capture-driven migration. In continuous notation, with \(\mathcal K_i=\dot C_{L,i}-\dot C_{J,i}-\ell_i\dot M_i\), \begin{equation}\label{eq:38}
\fitmath{\mathcal E_i=\int\mathcal K_i\,dt
       +\int(\ell_i-\ell_{i,0})\dot M_i\,dt.}
\end{equation} The second term includes the effect of prior evolution from all sources. We instead use every accepted step and its pre-step orbital state. Suppressing the planet index, let \(n\) label an accepted step from state \(n\) to state \(n+1\); \(\Delta\) denotes that step's increment. Then define \(\Delta C_{o,n}=\Delta C_{L,n}-\Delta C_{J,n}\), \(\Delta P_{o,n}=\Delta P_{L,n}-\Delta P_{J,n}\), and \(\Delta\mathcal K_n=\Delta C_{o,n}-\ell_n\Delta M_n\). The conservative update gives the exact identity \begin{equation}\label{eq:39}
\fitmath{\ell_{n+1}-\ell_n
 =\frac{\Delta\mathcal K_n}{M_{n+1}}
  +\frac{\Delta G_n}{M_{n+1}}
  +\frac{\Delta P_{o,n}}{M_{n+1}}.}
\end{equation} Table~\ref{tab:11} sums the three terms separately. All 11,912, 6,352, and 12,160 accepted steps are present for the lighter PDS, heavier PDS, and Solar-System histories. Steps duplicated by restarts agree and are counted once, and steps from abandoned run segments are excluded. The summed identities reproduce the measured changes to better than \(10^{-14}\) in code specific-angular-momentum units.

\begin{table*}[tp]
\centering
\caption{Accepted-step contributions to orbital specific angular momentum, in units of $10^{-3}\ell_{i,0}$ for each planet. Capture uses the pre-step orbital specific angular momentum and all impulses are divided by the post-step planetary mass. The three columns sum to the measured total before rounding. These are contributions to $\Delta\ell_i$, not percentages of radial displacement. Endpoints match Tables~\ref{tab:8}--\ref{tab:10}.}\label{tab:11}
\small\setlength{\tabcolsep}{2pt}
\begin{tabular}{@{}lllll@{}}
\toprule
\parbox[t]{0.2600\textwidth}{\raggedright Run and planet\strut} & \parbox[t]{0.1725\textwidth}{\centering Capture\strut} & \parbox[t]{0.1725\textwidth}{\centering Gravity\strut} & \parbox[t]{0.1725\textwidth}{\centering Pressure\strut} & \parbox[t]{0.1725\textwidth}{\centering Measured total\strut} \\[4pt]
\midrule
\parbox[t]{0.2600\textwidth}{\raggedright PDS \(2\,M_{\rm J}\), inner\strut} & \parbox[t]{0.1725\textwidth}{\centering +13.3\strut} & \parbox[t]{0.1725\textwidth}{\centering +13.1\strut} & \parbox[t]{0.1725\textwidth}{\centering -0.293\strut} & \parbox[t]{0.1725\textwidth}{\centering +26.0\strut} \\[4pt]
\parbox[t]{0.2600\textwidth}{\raggedright PDS \(2\,M_{\rm J}\), outer\strut} & \parbox[t]{0.1725\textwidth}{\centering -23.8\strut} & \parbox[t]{0.1725\textwidth}{\centering -18.2\strut} & \parbox[t]{0.1725\textwidth}{\centering +0.949\strut} & \parbox[t]{0.1725\textwidth}{\centering -41.1\strut} \\[4pt]
\parbox[t]{0.2600\textwidth}{\raggedright PDS \(6\,M_{\rm J}\), inner\strut} & \parbox[t]{0.1725\textwidth}{\centering +30.1\strut} & \parbox[t]{0.1725\textwidth}{\centering +13.3\strut} & \parbox[t]{0.1725\textwidth}{\centering +2.88\strut} & \parbox[t]{0.1725\textwidth}{\centering +46.3\strut} \\[4pt]
\parbox[t]{0.2600\textwidth}{\raggedright PDS \(6\,M_{\rm J}\), outer\strut} & \parbox[t]{0.1725\textwidth}{\centering -41.1\strut} & \parbox[t]{0.1725\textwidth}{\centering -19.8\strut} & \parbox[t]{0.1725\textwidth}{\centering +0.672\strut} & \parbox[t]{0.1725\textwidth}{\centering -60.2\strut} \\[4pt]
\parbox[t]{0.2600\textwidth}{\raggedright Solar \(2\,M_{\rm J}\), inner\strut} & \parbox[t]{0.1725\textwidth}{\centering +26.4\strut} & \parbox[t]{0.1725\textwidth}{\centering +12.2\strut} & \parbox[t]{0.1725\textwidth}{\centering -0.755\strut} & \parbox[t]{0.1725\textwidth}{\centering +37.8\strut} \\[4pt]
\parbox[t]{0.2600\textwidth}{\raggedright Solar \(2\,M_{\rm J}\), outer\strut} & \parbox[t]{0.1725\textwidth}{\centering -59.8\strut} & \parbox[t]{0.1725\textwidth}{\centering -22.3\strut} & \parbox[t]{0.1725\textwidth}{\centering +1.51\strut} & \parbox[t]{0.1725\textwidth}{\centering -80.5\strut} \\[4pt]
\bottomrule
\end{tabular}
\end{table*}

Capture contributes 51.0\% and 58.0\% of the net specific-angular-momentum changes in the lighter PDS disk, 65.0\% and 68.2\% in the heavier disk, and 69.8\% and 74.2\% in the Solar-System run, inner then outer. It is the larger contribution in each case, but capture and gravity are nearly equal for the lighter PDS inner planet. The pressure term reinforces convergence for the heavier PDS inner planet and opposes it for the other five cases. These are additive contributions to \(\Delta\ell\), not percentages of radial displacement or predictions with any force switched off. The full sum recovers the circular radius through \(a=(\ell^2/GM_\star)\). Both capture and torques must therefore be retained when using the inflow budget to predict migration and future feeding.

\section{Discussion: testing the delivery theory}\label{sec:6}

The scaled configuration can be viewed as another planetary system with its own central mass, planets, and disk. A delivery theory must predict the coupled evolution of that system. Preserving the equality between the Hill radius \(R_H\) of the planetary seeds and gas scale height \(H\) retains the three-dimensional accretion problem while making substantial reservoir evolution less expensive to compute. The aim is to test and constrain accretion
theory efficiently in the scaled setting and then apply it to the unscaled system.

The calculations in this study measure delivery to an absorbing Hill region, not accretion onto a planetary surface. Retained gas contributes to the planetary mass and angular-momentum ledger, without resolved CPD transport. The orbits are constrained to remain circular, and direct planet--planet gravity is omitted. Separately, we retain Scaling B, periodic-sector compression, as a comparison of scaling approaches, but defer its detailed physical and numerical evaluation to a later paper.

The scaled--unscaled comparison is encouraging. At the 1\% inventory target, Scaling A runs 9.31 times faster than the unscaled case while accreting 29\% less gas. The inner planet receives 35.4\% rather than 31.6\% (Table~\ref{tab:4}). Both systems favor delivery to the outer planet. Obtaining this similar early mass sharing at roughly one ninth of the computational cost strengthens the case for using scaled systems to develop and test the delivery theory.

This methods paper uses coarse meshes to obtain exploratory results quickly. That practical choice is independent of the scaling. The matched benchmark compares approximately equal local physical resolution, so its measured speedup does not come from coarsening the scaled calculation. The numerical setup reports the resolution of each simulation.

Orbital advection is complementary to our scaling. The physical acceleration and the equal-resolution grid saving of Section~\ref{sec:2.5} remain available when both systems use orbital advection, while the timestep ratio and cost per cell-step depend on the numerical scheme. The relevant comparison is therefore the scaled system with orbital advection against the unscaled system with orbital advection. Without advection, orbital speeds, which Scaling A reduces by \(\Gamma\) together with the ceiling, can add up to a further factor \(\Gamma\) where they control the CFL condition. Advection removes the background orbital speed from the azimuthal constraint, moving orbit-controlled limits toward the acoustic case, which carries no compression penalty. Under the nominal duration and grid-cost assumptions, comparable acoustic limits give \(\Gamma^4\), while corresponding velocity-dominated ceiling limits can supply an additional factor approaching \(\Gamma\), since both the ceiling and the background speed scale as \(1/\Gamma\). These are conditional estimates, not bounds on the measured saving, as the residual flow and limiting cells can differ between the two systems. Scaling results will be specific to the system modelled because the effective compression changes during the course of run.

\subsection{Completed and remaining comparisons}\label{sec:6.1}

The two PDS calculations (Sections~\ref{sec:4.6}--\ref{sec:4.7}) supply the gas-inventory, mass-partition, orbital, and height-dependent delivery measurements. The \(\Gamma=3\) Solar-System calculation (Section~\ref{sec:4.8}) adds a second planetary architecture with substantial growth and convergent migration, and the eight-rank benchmark (Section~\ref{sec:4.5}) separates the computational saving into physical duration, mean step cost, and timestep behavior. The other Solar-System compression choices are scaling estimates, not simulations. We defer simulation of the unscaled Solar-System due to its prohibitive compute budget. Development and predictive validation of the delivery closure belong to the companion theory paper.

\section{Conclusions}\label{sec:7}

\textbf{Scaling.} Scaling A provides two distinct economies. At a fixed \(R_H=H\) starting regime, the fixed-coefficient torque budget predicts a physical clearing-time reduction by \(\Gamma^2\). The contracted domain separately saves \(\Gamma^2\) cells at equal local resolution. Together these give a nominal work saving \(\Gamma^4\chi_1\), where \(\chi_1\) is the scaled-to-reference mean-timestep ratio (Section~\ref{sec:2.5}). Any contribution from the adopted velocity ceiling is prescription dependent. Scaling A does not preserve the Toomre parameter at fixed gas mass, because the increased surface density lowers \(Q_{\rm K}\). Because gas self-gravity is absent, the Toomre criterion is not a stability or stopping condition for these simulations, but \(Q\) remains an important physical diagnostic when applying the results to physical systems or adding gravitational transport.

\textbf{Matched benchmark.} At approximately matched local physical resolution, Scaling A delivers comparable captured masses and division of gas between the planets with a measured 9.31-fold computational speedup. The eight-rank comparison reaches the 1\% inventory-reduction target in 66.2 minutes unscaled and 7.12 minutes with Scaling A at \(\Gamma=1.75\). The shorter physical duration and larger mean timestep supply the saving; the measured mean step cost is slightly higher in the scaled run. The accretion comparison is encouraging in light of this large saving. Scaling A accretes 29\% less gas, but the inner planet's share differs by only 3.8 percentage points, with the outer planet receiving most of the gas in both cases. This combination of similar early mass sharing and roughly ninefold computational economy makes the scaled system a useful setting for testing accretion theory.

\textbf{Scaled PDS 70.} The two scaled PDS 70 simulations demonstrate substantial accretion and reservoir evolution in less than a day each. The analyzed moving-orbit inventory reductions are 50.2\% and 56.5\%, whereas the fixed-initial-orbit reductions are 32.6\% and 22.5\%. Migration therefore contributes substantially to the changing feeding inventory. In the heavier disk, subsequent capture reaches \(0.609\,M_{\rm J}\) and the planetary separation nearly halves. Across the two runs, 37.4--45.5\% of subsequent captured mass enters above half a Hill radius, but it carries only 24.0--28.5\% of the signed internal capture angular momentum. The angular-momentum budgets retain both internal CPD and orbital delivery. Step-weighted capture supplies 51--68\% of the orbital-specific-angular-momentum changes, with a nearly equal capture--gravity split for the lighter inner planet. The total budgets close to approximately two parts per billion.

\textbf{Scaled Solar System.} At \(\Gamma=3\), Scaling A predicts an 81-fold computational saving---nearly two orders of magnitude---at equal local resolution and unchanged timestep, combining ninefold faster torque-driven clearing with a ninefold reduction in cell count. The first scaled Solar-System run uses \(\Gamma=3\) and a \(2\,M_{\rm J}\) reservoir, with initial \(Q_{{\rm K},\min}\simeq1.34\). It reduces the moving-orbit interplanetary inventory by approximately half in 13.6 hours on eight MPI ranks. Its planets grow to \(0.0617\) and \(0.171\,M_{\rm J}\), and their separation contracts by 43.6\%. The outer planet receives 75\% of the gas accreted during evolution, excluding gas already inside the Hill spheres at initialization. The step-weighted capture contribution accounts for 69.8\% and 74.2\% of the inner and outer changes in orbital specific angular momentum; gravity reinforces convergence and pressure weakly opposes it. The buffered sink-patch geometry, rather than a self-gravity condition, ends the calculation. Within the fixed initial orbital interval the estimated gas reduction is only about 14\%, again distinguishing redistribution from migration-induced changes in the feeding inventory. The final values \(R_H/H=1.28\) and 1.50 imply \(q_{\rm th}=6.26\) and 10.1. The calculation therefore samples growth from \(R_H=H\) into a changed feeding geometry, providing a test of how delivery responds as the Hill sphere extends farther above the disk scale height.

\textbf{Outlook.} Together with the extended simulations, the benchmark demonstrates both the computational advantage and practical access to substantial evolution. A specific aim is to constrain the joint delivery of mass and specific angular momentum, since those exchanges couple CPD supply, migration, and the gas available for further growth. This approach allows the gas accretion theory to be tested in faster scaled systems before application to the original planetary systems.

\FloatBarrier

\begin{acknowledgments}\renewcommand{\makeLineNumber}{}
\section*{Acknowledgments}
This manuscript was prepared in collaboration with GPT-6 Pro and GPT-6 Astra High, with editorial assistance from GLM-5.3 and Opus 5.5.
\end{acknowledgments}

\software{Athena++ \citep{ref7}, NumPy \citep{ref61}, Matplotlib \citep{ref62}.}

\appendix

\setcounter{table}{0}\renewcommand{\thetable}{A\arabic{table}}\renewcommand{\theHtable}{A\arabic{table}}

\section{Numerical setup and timestep diagnostics}\label{sec:A}

\subsection{Initial conditions and numerical prescriptions}\label{sec:A.1}

For the scaled PDS calculations, the units are \(GM_\star=1\), length 19.7 au, mass \(0.142\,M_\odot\), and time 37.0 yr. One Jupiter mass is 0.00673 code mass. Both hemispheres are counted once by doubling the modeled upper half.

Let \(W(R)\) be the radial taper, with inner and outer centers \(R_{\rm in}\) and \(R_{\rm out}\), width \(w_{\rm tap}\), and surface-density normalization \(\Sigma_0\). In cylindrical coordinates \((R,z)\), the initializer is \begin{equation}\label{eq:40}
\fitmath{\begin{aligned}
 W(R)&=\tfrac14[1+\tanh((R-R_{\rm in})/w_{\rm tap})]\\
     &\quad\times[1-\tanh((R-R_{\rm out})/w_{\rm tap})],\\
 \rho(R,z)&=\frac{\Sigma_0[W(R)+10^{-4}]}{\sqrt{2\pi}H(R)}\\
 &\quad\times\exp\!\left[\frac{GM_\star}{c_s^2}
 \left(\frac{1}{\sqrt{R^2+z^2}}-\frac1R\right)\right].
 \end{aligned}}
\end{equation} Here \(H(R)=c_s/\sqrt{GM_\star/R^3}\), \(w_{\rm tap}=1\) au, and the taper centers are the input orbits, 12.6 and 19.7 au. The code normalization \(\Sigma_0=0.00715\) gives \(2\,M_{\rm J}\) before initial capture on the native grid. Initial radial and vertical velocities vanish. An advected passive tracer distinguishes the tapered reservoir from the tenuous background. Its concentration is initialized as \(C=\rho_{\rm res}/(\rho_{\rm res}+\rho_{\rm bg})
=W/(W+10^{-4})\), where \(\rho_{\rm res}\) and \(\rho_{\rm bg}\) are the two density contributions above. The conserved tracer density is \(\rho C\). It has no dynamical feedback and tags reservoir material, not membership in the initial interplanetary interval. The pressure-supported azimuthal velocity \(v_\phi\) is initialized using centered pressure differences. Writing \(W'(R)=dW/dR\) for a radial derivative, not a scaled quantity, its analytic form is \begin{equation}\label{eq:41}
\fitmath{v_\phi^2=\frac{GM_\star}{R}
  +c_s^2\left[\frac{RW'(R)}{W(R)+10^{-4}}-\frac32\right].}
\end{equation} Both planetary potentials are present without a mass ramp. Their initial azimuths are \(\pi/2\) (inner) and \(\pi\) (outer). The physical softening length is 0.354 au. For displacement vector \(\mathbf d\) from planet to gas and \(d=|\mathbf d|\), the planetary acceleration is \(-GM_i\mathbf d/d^3\) outside this length \(\epsilon\), and \(GM_i\mathbf d(3d/\epsilon-4)/\epsilon^3\) inside it. The upper polar boundary and midplane are reflecting; radial boundaries admit outflow only and azimuth is periodic. The sound speed is 0.105 in code units (20 K), the density floor is \(10^{-12}\), and the component velocity ceiling is \(3\sqrt{2GM_\star/r_{\rm inner}}=6.71\). The isothermal equation of state has no additional temperature-reset operation.

The root grid, blocks, and single added adaptive level are specified in Section~\ref{sec:3.1}. The refinement neighborhood uses two Hill radii. Capture surfaces are faceted in the volume-coordinate representation; exterior cut volumes, conservative merging, and the circular-orbit update are coupled at each accepted step. Managed patches extend three finest cells beyond the surface bounds and require a native-cell gap between patches and unique finest-level ownership. VL2, second-order reconstruction, HLLE, and CFL 0.3 are retained, with the aggregate-patch reconstruction limitation stated in Section~\ref{sec:4.1}. No dynamics is evolved within the capture surface. Step budgets are retained every accepted step; radial and cumulative height/azimuth capture summaries are sampled every 20 steps and at segment endpoints. The latter preserve aggregate capture quantities without storing each individual capture event.

\subsection{CFL reconstruction and sampled limiting cells}\label{sec:A.2}

Let \(k\) label an exterior cell and \(d\) a radial, polar, or azimuthal direction. Write \(\Delta x_{k,d}\) for its physical cell width and \(v_{k,d}\) for the corresponding velocity component. With Courant safety factor \(C_{\rm CFL}=0.3\), the isothermal, nonmagnetic solver without orbital advection evaluates \begin{equation}\label{eq:42}
\fitmath{\Delta t_{\rm CFL}=C_{\rm CFL}
 \min_{k,d}\frac{\Delta x_{k,d}}{|v_{k,d}|+c_s}.}
\end{equation} The velocities include any component clipping; fast cells are not excluded from this minimum. A ceiling-controlled comparison requires the controlling components to reach their respective ceilings in both systems at comparable physical cell widths. Finding a capped cell in the scaled run alone does not establish that comparison. Motions on the preserved local scales \(H\) and \(R_H\) need not decrease with stellar orbital speed.

The diagnostic identifies the limiting cell's position, height \(|z|/H\), density, refinement level, physical width, velocity component, sound speed, and proximity to the ceiling. Here \(H\) is evaluated at the cell's cylindrical radius. These distinguish acoustic, orbital-motion, and fast atmospheric-flow limits. The reconstructed minimum is compared with the accepted timestep, allowing for separate timestep-growth or end-time restrictions. Checkpoint samples identify individual limiting states, but do not measure the fraction of steps controlled by each regime. That requires a per-step limiting-cell record.

At 20.4 simulated years, the limiting cell lies approximately \(4.5H\) above the midplane at the density floor. Its azimuthal speed is approximately \(50c_s\), or 78\% of the ceiling; velocity supplies about 98\% of the CFL denominator. Reconstruction reproduces the timestep to roundoff. This state is velocity-dominated but below the ceiling.

At cycle 826 (30.0 simulated years), the limiting radial component lies at \(4.26H\), also at the density floor. In code units, the density is \(10^{-12}\) and radial momentum density is \(-6.71\times10^{-12}\). Their ratio equals the negative component ceiling, \(-6.71\), to floating-point precision. The physical cell width is 0.0132 and sound speed 0.105 in code units, giving \(\Delta t=5.81\times10^{-4}\), consistent with the saved timestep to roundoff when evaluated with full-precision inputs. This checkpoint is ceiling-limited. Neither sample establishes continuous ceiling control or the mean-timestep gain in the matched benchmark.

\subsection{Reference and compressed geometries}\label{sec:A.3}

Table~\ref{tab:12} compares the reference and all-radii compressed geometries. These are nominal mapped quantities before discrete initial capture. The reference star is \(0.76\,M_\odot\); Scaling A uses \(0.76/1.75^3=0.142\,M_\odot\), with unchanged sound speed and nominal seed masses. Entries containing a pair are inner then outer. The reference and Scaling A are the two configurations of the 1\% benchmark.

\begin{table}[tp]
\centering
\caption{Nominal reference and Scaling A geometries at $\Gamma=1.75$. Paired entries are inner then outer; frequencies, heights, Hill radii, and ceiling speeds are ratios to the corresponding reference quantities. The initial Hill-to-height row gives $R_H/H$ itself, not a scaled-to-reference ratio. Values precede discrete initial capture. The boundary radii use the full stated reference domain.}\label{tab:12}
\small\setlength{\tabcolsep}{2pt}
\begin{tabular}{@{}lll@{}}
\toprule
\parbox[t]{0.4800\columnwidth}{\raggedright Quantity\strut} & \parbox[t]{0.2400\columnwidth}{\centering Reference\strut} & \parbox[t]{0.2400\columnwidth}{\centering Scaling A\strut} \\[4pt]
\midrule
\parbox[t]{0.4800\columnwidth}{\raggedright Orbital radii (au)\strut} & \parbox[t]{0.2400\columnwidth}{\centering 22.0, 34.5\strut} & \parbox[t]{0.2400\columnwidth}{\centering 12.6, 19.7\strut} \\[4pt]
\parbox[t]{0.4800\columnwidth}{\raggedright Radial domain (au)\strut} & \parbox[t]{0.2400\columnwidth}{\centering 13.8--46.6\strut} & \parbox[t]{0.2400\columnwidth}{\centering 7.89--26.6\strut} \\[4pt]
\parbox[t]{0.4800\columnwidth}{\raggedright Local \(\Omega'/\Omega\)\strut} & \parbox[t]{0.2400\columnwidth}{\centering 1, 1\strut} & \parbox[t]{0.2400\columnwidth}{\centering 1, 1\strut} \\[4pt]
\parbox[t]{0.4800\columnwidth}{\raggedright Local \(H'/H\)\strut} & \parbox[t]{0.2400\columnwidth}{\centering 1, 1\strut} & \parbox[t]{0.2400\columnwidth}{\centering 1, 1\strut} \\[4pt]
\parbox[t]{0.4800\columnwidth}{\raggedright Local \(R_H'/R_H\)\strut} & \parbox[t]{0.2400\columnwidth}{\centering 1, 1\strut} & \parbox[t]{0.2400\columnwidth}{\centering 1, 1\strut} \\[4pt]
\parbox[t]{0.4800\columnwidth}{\raggedright Initial \(R_H/H\)\strut} & \parbox[t]{0.2400\columnwidth}{\centering 1, 1\strut} & \parbox[t]{0.2400\columnwidth}{\centering 1, 1\strut} \\[4pt]
\parbox[t]{0.4800\columnwidth}{\raggedright Inner-wall \(\Omega_K'/\Omega_K\)\strut} & \parbox[t]{0.2400\columnwidth}{\centering 1\strut} & \parbox[t]{0.2400\columnwidth}{\centering 1\strut} \\[4pt]
\parbox[t]{0.4800\columnwidth}{\raggedright Ceiling \(v_{\rm cap}'/v_{\rm cap}\)\strut} & \parbox[t]{0.2400\columnwidth}{\centering 1\strut} & \parbox[t]{0.2400\columnwidth}{\centering 0.571\strut} \\[4pt]
\bottomrule
\end{tabular}
\end{table}

\subsection{Solar-System numerical specification}\label{sec:A.4}

The Solar-System calculation uses \(GM_\star=1\), length unit 3.19 au, time unit 4.72 yr, and central mass \(M_\star=M_\odot/27\). Table~\ref{tab:13} distinguishes its parameters from the PDS implementation. Both use the softened force law in Appendix~\ref{sec:A.1}, which is exactly Newtonian outside \(\epsilon\). The initial Solar-System ratios \(\epsilon/R_H=0.335\) and 0.136 are also the maximum values over all saved accepted steps. The nominal Hill capture surfaces therefore remain outside the softened region throughout the reported evolution.

\begin{table}[tp]
\centering
\caption{Numerical parameters for the simulations. Code quantities use each system's own units; identical code ceilings do not imply identical physical velocities. Both PDS reservoir masses use the same numerical prescriptions. The Solar-System softening is not the PDS physical softening.}\label{tab:13}
\small\setlength{\tabcolsep}{2pt}
\begin{tabular}{@{}lll@{}}
\toprule
\parbox[t]{0.4800\columnwidth}{\raggedright Numerical quantity\strut} & \parbox[t]{0.2400\columnwidth}{\centering PDS disks\strut} & \parbox[t]{0.2400\columnwidth}{\centering Solar-System disk\strut} \\[4pt]
\midrule
\parbox[t]{0.4800\columnwidth}{\raggedright Length unit (au)\strut} & \parbox[t]{0.2400\columnwidth}{\centering 19.7\strut} & \parbox[t]{0.2400\columnwidth}{\centering 3.19\strut} \\[4pt]
\parbox[t]{0.4800\columnwidth}{\raggedright Sound speed (code)\strut} & \parbox[t]{0.2400\columnwidth}{\centering 0.105\strut} & \parbox[t]{0.2400\columnwidth}{\centering 0.0828\strut} \\[4pt]
\parbox[t]{0.4800\columnwidth}{\raggedright Sound speed (km s\(^{-1}\))\strut} & \parbox[t]{0.2400\columnwidth}{\centering 0.266\strut} & \parbox[t]{0.2400\columnwidth}{\centering 0.266\strut} \\[4pt]
\parbox[t]{0.4800\columnwidth}{\raggedright Softening \(\epsilon\) (code)\strut} & \parbox[t]{0.2400\columnwidth}{\centering 0.0179\strut} & \parbox[t]{0.2400\columnwidth}{\centering 0.0111\strut} \\[4pt]
\parbox[t]{0.4800\columnwidth}{\raggedright Softening \(\epsilon\) (au)\strut} & \parbox[t]{0.2400\columnwidth}{\centering 0.354\strut} & \parbox[t]{0.2400\columnwidth}{\centering 0.0354\strut} \\[4pt]
\parbox[t]{0.4800\columnwidth}{\raggedright Density floor (code)\strut} & \parbox[t]{0.2400\columnwidth}{\centering \(10^{-12}\)\strut} & \parbox[t]{0.2400\columnwidth}{\centering \(10^{-12}\)\strut} \\[4pt]
\parbox[t]{0.4800\columnwidth}{\raggedright Velocity ceiling (code)\strut} & \parbox[t]{0.2400\columnwidth}{\centering 6.71\strut} & \parbox[t]{0.2400\columnwidth}{\centering 6.71\strut} \\[4pt]
\parbox[t]{0.4800\columnwidth}{\raggedright Velocity ceiling (km s\(^{-1}\))\strut} & \parbox[t]{0.2400\columnwidth}{\centering 16.9\strut} & \parbox[t]{0.2400\columnwidth}{\centering 21.5\strut} \\[4pt]
\parbox[t]{0.4800\columnwidth}{\raggedright Upper wedge (rad)\strut} & \parbox[t]{0.2400\columnwidth}{\centering 0.405\strut} & \parbox[t]{0.2400\columnwidth}{\centering 0.325\strut} \\[4pt]
\parbox[t]{0.4800\columnwidth}{\raggedright Root mesh \((r,\theta,\phi)\)\strut} & \parbox[t]{0.2400\columnwidth}{\centering \(36\times6\times112\)\strut} & \parbox[t]{0.2400\columnwidth}{\centering \(36\times6\times112\)\strut} \\[4pt]
\parbox[t]{0.4800\columnwidth}{\raggedright Additional refinement levels\strut} & \parbox[t]{0.2400\columnwidth}{\centering 1\strut} & \parbox[t]{0.2400\columnwidth}{\centering 1\strut} \\[4pt]
\bottomrule
\end{tabular}
\end{table}

The Solar-System run uses VL2, second-order reconstruction, HLLE, and CFL 0.3, with reflecting polar, outflow-only radial, and periodic azimuthal boundaries. The code floor corresponds to approximately \(6.76\times10^{-22}\,\mathrm{g\,cm^{-3}}\) in its own units. The recorded solver base revision is \texttt{630f8de3}. Continuation from cycle 10,843 uses a local refinement-coverage correction that includes the padded moving capture patches and predicted sweep; it does not add a refinement level or change the capture rule. The two executable SHA-256 identifiers begin \texttt{22be47b95985619e} and \texttt{206cc6db00ba4893}, respectively. These identify the two builds without implying that the base commit alone includes the continuation correction.

\subsection{Run records}\label{sec:A.5}

\textbf{Matched benchmark.} Both eight-rank calculations in Section~\ref{sec:4.5} ran on the same Apple Mac mini (M4 Pro, 14 CPU cores: 10 performance and four efficiency cores; 48 GB unified memory). Each used eight MPI processes with one thread per process; no explicit core-affinity option was supplied. The solver was compiled with Apple Clang through the \texttt{mpicxx} wrapper, using \texttt{-O3\ -std=c++11}, with MPI enabled and OpenMP disabled. The MPI implementation was Homebrew Open MPI 5.0.10. Both runs used the same modified Athena++ executable, whose SHA-256 identifier begins \texttt{206cc6db00ba4893}, based on solver revision \texttt{630f8de3} with the local capture and refinement-coverage changes described above. This is the same build used for the Solar-System continuation in Appendix~\ref{sec:A.4}. The executable identifier, rather than the base revision alone, distinguishes this build. The saved Makefile and compiler-configuration hashes agree between the two runs. The common controller used 20-step segments and the same checkpoint and fractional-volume inventory checks; the reported wall time includes those operations.

\textbf{Extended PDS run.} The \(2\,M_{\rm J}\) scaled PDS calculation of Section~\ref{sec:4.6} ran in an original allocation of 16.3 hours on 14 MPI ranks, followed by a 1.32-hour continuation on eight ranks: 17.6 hours, or about 239 rank-hours, in total. This total includes segment startup, checkpointing, endpoint checks, and the stopped tail of the original allocation with its replay from the saved checkpoint; it excludes separate diagnostic investigations. The continuation's history replaces the replayed tail, so no interval or capture increment is counted twice. The endpoint at cycle 11,912 has matched radial profiles, reservoir diagnostics, capture summaries, and a full restart checkpoint. The continuation reports finite passive-tracer bound excursions without clipping the tracer or altering the gas update; nonpositive gas mass and nonfinite states remain fatal.

\textbf{Solar-System run.} The 13.6 hours of production execution for the \(\Gamma=3\) Solar-System calculation of Section~\ref{sec:4.8} include the original allocation and its continuation, including the short replayed tail, but exclude intervening downtime and separate diagnostic work; other workloads were present. The last full restart checkpoint is at cycle 11,558 (28.8 yr); Figure~\ref{fig:6} uses the later saved diagnostic histories rather than an assumed checkpoint at the plotted endpoint. The continuation included a refinement-coverage correction that protects the existing managed patches without adding a refinement level (Appendix~\ref{sec:A.4}).
\FloatBarrier
\raggedbottom

\end{document}